\documentclass[12pt,onecolumn,draftclsnofoot]{IEEEtran}

\newtheorem{thm}{Theorem}

\newtheorem{lem}{Lemma}

\newtheorem{defn}{Definition}
\newtheorem{remark}{Remark}

\usepackage[final]{graphicx}
\usepackage[reqno]{amsmath}
\usepackage{amssymb}
\usepackage{amsmath}
\usepackage{subfig}
\usepackage{epstopdf}
\usepackage{xcolor}
\usepackage{float}
\usepackage[paper=letterpaper,tmargin=1in,bmargin=.8in,left=.8in, right=.8in]{geometry}
\usepackage{comment}

\newtheorem{lemma}{Lemma}
\newtheorem{corollary}{Corollary}

\begin{document}

\sloppy
 
\title{Joint Uplink-Downlink Cooperative Interference Management with Flexible Cell Associations}
\author{{\large{Manik Singhal, {\em Student Member, IEEE}, Tolunay Seyfi, {\em Student Member, IEEE}, and Aly El Gamal, {\em Senior Member, IEEE}}.}
\thanks{Manik Singhal, Tolunay Seyfi, and Aly El Gamal are with the ECE Department at Purdue University, West Lafayette, IN (e-mail: msingha,tseyfi,elgamala@purdue.edu).}
\thanks{This paper was presented in part at the 2016 and 2018 versions of the International Symposium on Information Theory (ISIT)~\cite{ElGamal-ISIT16, Singhal-ElGamal-ISIT18}, and the 2017 version of the Allerton Conference on Signals, Systems and Computers~\cite{Singhal-ElGamal-Allerton17}}
}
\maketitle

\begin{abstract}
We study information theoretic models of interference networks that consist of $K$ Base Station (BS) - Mobile Terminal (MT) pairs. Each BS is connected to the MT carrying the same index as well as $L$ following MTs. We fix the value of $L$ and study the per user Degrees of Freedom (puDoF) in large networks. We assume that each MT can be associated with $N_c$ BSs, and these associations are determined by a cloud-based controller that has a global view of the network. An MT has to be associated with a BS, for the BS to transmit its message in the downlink, or have its decoded message in the uplink. We propose puDoF inner bounds for arbitrary values of $L$ when only the uplink is considered, and characterize the uplink puDoF value when only zero-forcing schemes are allowed. We then introduce new achievable average uplink-downlink puDoF values, and show their optimality for the range when $N_c \leq \frac{L}{2}$ and when we restrict our attention to zero-forcing schemes. Additionally, for the remaining range, we characterize the optimal downlink scheme when the uplink-optimal associations are used. Finally, we show that the proposed scheme is information theoretically optimal for Wyner's linear interference network.
\end{abstract}
\section{Introduction}
The fifth generation of cellular networks is expected to bring new paradigms to wireless communications, that exploit recent technological advancements like cloud computing and cooperative communication (also known as Coordinated Multi-Point or CoMP). In particular, the rising interest in Cloud Radio Access Networks (C-RAN) (see e.g.,~\cite{CRAN}-\cite{CRAN-Simeone-2}) holds a promise for such new paradigms. These paradigms require new theoretic frameworks to identify fundamental limits and suggest insights that are backed by rigorous analysis. The focus of this work is to identify associations between cell edge mobile terminals and base stations, that maximize the average rate across both uplink and downlink sessions, while allowing for associating one mobile terminal with more than one base station and using cooperative transmission and reception schemes between base stations in the downlink and uplink sessions, respectively. With a cloud-based controller, optimal decisions for these associations can take into account the whole network topology, with the goal of maximizing a sum rate function. 

Cloud-based CoMP communication is a promising new technology that could significantly enhance the rates of cell edge users (see~\cite{Veeravalli-ElGamal-Cambridge} and~\cite{CoMP-book} for an overview of CoMP). In~\cite{Annapureddy-ElGamal-Veeravalli-IT12}, an information theoretic model was studied where cooperation was allowed between transmitters, as well as between receivers (CoMP transmission and reception). CoMP transmission and reception schemes in cellular networks are applicable in the downlink and uplink, respectively. The model in~\cite{Annapureddy-ElGamal-Veeravalli-IT12} assumed that each message can be available at $M_t$ transmitters and can be decoded through $M_r$ received signals. It was shown that full Degrees of Freedom (DoF) can be achieved if $M_t+M_r \geq K+1$, where $K$ is the number of transmitter-receiver pairs (users) in the network.  

Recently in~\cite{Ntranos-arXiv14}, alternative frameworks for cooperation in both downlink and uplink were introduced. The new frameworks are based on the concept of \emph{message passing} between base stations. In the downlink, quantized versions of the analog transmit signals are being shared between base station transmitters. The supporting key idea is that information about multiple messages can be shared from one transmitter to another with the cost of sharing only one whole message (of the order of $\log P$, where $P$ is the transmit power), if we only share information needed to cancel the interference caused by the messages at unintended receivers, through dirty paper coding (see~\cite{DPC}). In the uplink, decoded messages are shared from one base station receiver to another, where they are used to cancel interference. It was shown in~\cite{Ntranos-arXiv14} that there is a duality in this framework between schemes that are used in the downlink and those that are used for the uplink, with the clear advantage that the same backhaul infrastructure can be used to support both scenarios. 
%

It is worth noting that the CoMP reception scheme introduced in~\cite{Annapureddy-ElGamal-Veeravalli-IT12} requires sharing of analog received signals over the backhaul. Also, the cooperative transmission through dirty paper coding introduced in~\cite{Ntranos-arXiv14} requires sharing of quantized analog signals over the backhaul, and also incurs a delay that scales with the size of the network. In this work, we consider sharing of only digital message information over the backhaul for both the downlink and uplink. Only whole messages could be shared over the backhaul, and hence, as special cases of our setting, we get the CoMP transmission paradigm of~\cite{Annapureddy-ElGamal-Veeravalli-IT12} and the message passing decoding paradigm of~\cite{Ntranos-arXiv14}. We first characterize the puDoF of message passing decoding in the uplink of locally connected interference networks. We then consider the problem of jointly optimizing the assignment of messages over the backhaul to maximize the average puDoF across both downlink and uplink sessions. We assume that each base station can be associated with $N_c$ mobile terminals, and that an association is needed whenever a mobile terminal's message is \emph{used} by a base station in either the downlink or the uplink. This usage of a message could be either for delivering the message in downlink, decoding the message in uplink, or for interference cancellation. We first show how our result for the uplink settles the average puDoF problem when $N_c \leq \frac{L}{2}$. We then tackle this problem when $N_c > L$, by fixing the uplink scheme to the optimal uplink-only scheme, that associates each mobile terminal with the $L+1$ base stations connected to it, and characterize the optimal downlink scheme under this constraint. The intuition behind this step is that full DoF is achieved in the uplink when $N_c > L$ through associating each mobile terminal with all $L+1$ base stations connected to it: Any change in that cell association is expected to decrease the uplink puDoF with a factor greater than the gain achieved for the downlink puDoF. We finally demonstrate the information theoretic optimality of the presented scheme for the linear interference network introduced by Wyner~\cite{Wyner} (when $L=1$). 

It is important to note that the assumptions are used to define a tractable problem whose solution can lead to constructive insights. For example, it was shown in~\cite{ElGamal-Annapureddy-Veeravalli-IT14} that imposing a downlink backhaul constraint where each message can be available at a specified maximum number of transmitters (maximum transmit set size constraint), can lead to solutions that are also useful to solve the more difficult and more relevant to practice problem, where an average transmit set size constraint is used instead of the maximum. Also, in~\cite{Bande-ElGamal-Veeravalli-arXiv16}, it was shown that solutions obtained for the locally connected network models, that are considered in this work, can be used to obtain solutions for the more practical cellular network models, by viewing the cellular network as a set of interfering locally connected subnetworks and designing a fractional reuse scheme that avoids interference across subnetworks.


\subsection{Related work}
In ~\cite{Zhiguo-Ding}~-~\cite{Gaurav-Nigam}, new stochastic geometry-based models for understanding base station cooperation in heterogeneous cellular networks are presented. Unlike the considered work, the base station locations are drawn from a homogeneous Poisson point process, instead of a fixed network topology model. In these works, a set of randomly located base stations jointly transmit and receive data, for decreasing intercell interference and improving coverage.

In ~\cite{Osvaldo-Simeone}, the uplink of an infrastructural network was modeled according to a standard linear Wyner-type model, which assumes both local connectivity and finite-capacity backhaul links for decoding at the base stations. Achievable rates were presented for basic successive-decoding-based cooperation strategies. Furthermore the influence of local and finite-capacity inter-BS connections on the achievable performance was studied.

In \cite{Xiang-Li}, a duality of IA schemes for both the uplink and the downlink transmissions was proposed for cellular relay networks. The goal was to find an inter-cell IA solution  then to design a zero-forcing filter for each BS to suppress intra-cell interference. A relation between the power allocation schemes for the uplink and the downlink transmissions was derived such that the same sum rate can be achieved with a sum transmit power constraint.

The distinguishing aspects of this work are studying cell association decisions that maximize the average rate over both uplink and downlink sessions under a limited backhaul budget, and considering simple and closer-to-practice cooperative zero-forcing transmission and message passing decoding reception schemes in the dowlink and uplink, respectively.


\section{System Model and Notation}\label{sec:model}
For each of the downlink and uplink sessions, we use the standard model for the $K-$user interference channel with single-antenna transmitters and receivers,
\begin{equation}
Y_i(t) = \sum_{j=1}^{K} H_{i,j}(t) X_j(t) + Z_i(t),
\end{equation}
where $t$ is the time index, $X_j(t)$ is the transmitted signal of transmitter $j$, $Y_i(t)$ is the received signal at receiver $i$, $Z_i(t)$ is the zero mean unit variance Gaussian noise at receiver $i$, and $H_{i,j}(t)$ is the channel coefficient from transmitter $j$ to receiver $i$ over time slot $t$. We remove the time index in the rest of the paper for brevity unless it is needed. The signals $Y_i$ and $X_i$ correspond to the receive and transmit signals at the $i^{\textrm{th}}$ base station and mobile terminal in the uplink, respectively, and the $i^{\textrm{th}}$ mobile terminal and base station in the downlink, respectively. For consistency of notation, we will always refer to $H_{i,j}$ as the channel coefficient between mobile terminal $i$ and base station $j$. Finally, we use $MT$ $i$ to denote mobile terminal $i$, and $BS$ $j$ to denote base station $j$.

\subsection{Channel Model}\label{sec:channel}
We consider the following locally connected interference network. The mobile terminal with index $i$ is connected to base stations $\{i,i-1,\cdots,i-L\}$, except the first $L$ mobile terminals, which are connected only to all the base stations with a similar or lower index. More precisely, 

\begin{equation}\label{eq:channel}
H_{i,j} = 0 \text { iff } i \notin \{j,j+1,\cdots,j+L\},\forall i,j \in [K],
\end{equation}
and all non-zero channel coefficients are drawn from a continuous joint distribution. We also assume that global channel state information is available at all mobile terminals and base stations.

Finally, we define the \textbf{interference set} to denote the set of receivers a transmitter is connected to.
\begin{defn}
 In the uplink, the interference set of MT $\alpha$ is the set of base stations with indices in the set $\{\alpha, \alpha -1, \cdots ,\alpha - L\}$.
 In the downlink, the Interference set of BS $\beta$ is the set of mobile terminals with indices in the set $\{\beta, \beta + 1, \cdots, \beta +L\}$.
\end{defn}
\subsection{Cell Association}
For each $i \in [K]$, let ${\cal C}_i \subseteq [K]$ be the set of base stations, with which mobile terminal $i$ is associated, i.e., those base stations that carry the terminal's message in the downlink and can have its decoded message for the uplink. Any subset of the transmitters in ${\cal C}_i$ may cooperatively transmit the message (word) $W_i$ to mobile terminal $i$ in the downlink. In the uplink, one of the base station receivers in ${\cal C}_i$ may decode $W_i$ and pass it to other receivers in the set. We consider a cell association constraint that bounds the cardinality of the set ${\cal C}_i$ by a number $N_c$; this constraint is one way to capture a limited backhaul capacity constraint where not all messages can be exchanged over the backhaul. 
\begin{equation}\label{eq:backhaul_constraint}
|{\cal C}_i| \leq N_c, \forall i\in [K].
\end{equation}

We would like to stress on the fact that we \textbf{only allow full messages to be shared over the backhaul}. More specifically, splitting messages into parts and sharing them as in~\cite{Wigger}, or sharing of quantized signals as in~\cite{Ntranos-arXiv14} is not allowed. 

We also make the following definitions for cell associations that \emph{cover} each mobile terminal with all base stations connected to it.

\begin{defn}
 We say that the cell association scheme is a \textbf{Full coverage association} if each mobile terminal is associated with all the base stations connected to it. More precisely, $ \forall{i} \in [K], \{i, i-1, \cdots, i-L\} \subseteq {\cal C}_i$.
\end{defn}

As shown later, full coverage associations lead to complete uplink interference cancellation.
\subsection{Degrees of Freedom}
Let $P$ be the average transmit power constraint at each transmitter, and let ${\cal W}_i$ denote the alphabet for message $W_i$. Then the rates $R_i(P) = \frac{\log|{\cal W}_i|}{n}$ are achievable if the decoding error probabilities of all messages can be simultaneously made arbitrarily small for a large enough coding block length $n$, and this holds for almost all channel realizations. The degrees of freedom $d_i, i\in[K],$ are defined as $d_i=\lim_{P \rightarrow \infty} \frac{R_i(P)}{\log P}$. The DoF region ${\cal D}$ is the closure of the set of all achievable DoF tuples. The total number of degrees of freedom ($\eta$) is the maximum value of the sum of the achievable degrees of freedom, $\eta=\max_{\cal D} \sum_{i \in [K]} d_i$.

For a $K$-user locally connected with connectivity parameter $L$, we define $\eta(K,L,N_c)$ as the best achievable $\eta$ on average taken over both downlink and uplink sessions over all choices of cell associations satisfying the backhaul load constraint in \eqref{eq:backhaul_constraint}. 
In order to simplify our analysis, we define the asymptotic per user DoF (puDoF) $\tau(L,N_c)$ to measure how $\eta(K,L,N_c)$ scales with $K$ while all other parameters are fixed,
\begin{equation}
\tau(L,N_c) = \lim_{K\rightarrow \infty} \frac{\eta(K,L,N_c)}{K}.
\end{equation}

We further define $\tau_D (L,N_c)$ and $\tau_U (L,N_c)$ as the puDoF when we optimize only for the downlink and uplink session, respectively.

\subsection{Interference Avoidance Schemes} 
We consider in this work the class of \textit{interference avoidance} schemes, where every receiver is either active or inactive. An active receiver can observe its desired signal with no interference. In the downlink, we are considering cooperative zero-forcing transmission where a message's interference is cancelled \emph{over the air} through cooperating transmitters. More precisely, for any zero-forcing scheme, the transmit signal at the $j^{\mathrm{th}}$ transmitter is given by,
\begin{equation}\label{eq:linear-precoding}
X_j = \sum_{i: j \in {\cal C}_i} X_{j,i},
\end{equation}
where $X_{j,i}$ depends only on message $W_i$. Further, each message is either not transmitted or allocated one degree of freedom. More precisely, let $\tilde{Y}_j=Y_j-Z_j, \forall j\in[K]$. Then, in addition to the constraint in~\eqref{eq:linear-precoding}, it is either case that the mutual information $I(\tilde{Y}_j;W_j)=0$ or it is the case that $W_j$ completely determines $\tilde{Y}_j$. Note that $\tilde{Y}_j$ can be determined from $W_j$ for the case where user $j$ enjoys interference-free communication, and $I(\tilde{Y}_j;W_j)=0$ for the other case where $W_j$ is not transmitted. We say that the $j^{\mathrm{th}}$ receiver is \emph{active} if and only if $I(\tilde{Y}_j;W_j)>0$. If the $j^{\mathrm{th}}$ receiver is active when using zero-forcing transmit beamforming, then $I(Y_i;W_i)=0, \forall i \neq j$. Finally, we say that the $j^{\mathrm{th}}$ transmitter is \emph{active} if $I\left(X_j;\{W_i: j\in {\cal C}_i\}\right)>0$.

In the uplink, we are considering zero-forcing of interference through message passing decoding, where a decoded message is passed through a cooperating receiver to other receivers wishing to remove the message's interference.
More precisely, we say that the $j^{\mathrm{th}}$ mobile terminal transmitter is active in the uplink if $I(X_j;W_j)>0$. Further, each active mobile terminal uses an optimal AWGN point-to-point code (see e.g.,~\cite{Cover-Thomas}) with transmit power $P$. For each active base station receiver with index $i$, if we denote the set of all messages that BS $i$ receives over the backhaul by ${B}_{i}$, then it has to be the case that there exists $j$ such that BS $i$ is associated with MT $j$ and $W_{j}$ determines $\tilde{Y_{i}}$, given perfect estimates of the messages shared in $B_i$. In other words, $\exists ~j$ s.t. $i \in {\cal C}_{j}$ and $I(\tilde{Y_{i}}; W_{j}|B_{i})>0$.
Further, $\forall k\in [K]: k \neq j$, $I(\tilde{Y_{i}^n}; W_{k}|B_i)\overset{n \rightarrow \infty}{\longrightarrow}0$, where $n$ is the block length. Note that the decoded message estimates become \emph{perfect} as the block length goes to infinity. We then make the following definition. 
\begin{defn}
 For an uplink zero-forcing scheme, we say that the MT-BS pair $\left(\text{MT } j, \text{BS } i\right)$ is a \textbf{decoding pair} if $W_j$ is decoded at base station $i$. More precisely, $i \in {\cal C}_{j}$, $I(\tilde{Y_{i}}; W_{j}|B_{i})>0$, and
$\forall k\in [K]: k \neq j$, $I(\tilde{Y_{i}^n}; W_{k}|B_i)\overset{n \rightarrow \infty}{\longrightarrow}0$.
\end{defn}


We add the superscript {\bf zf} to the puDoF symbol when we impose the constraint that the coding scheme that can be used has to be a zero-forcing scheme. For example, $\tau_U^{\textrm{zf}}(L,N_c)$ denotes the puDoF value when considering only the uplink and impose the restriction to message passing decoding zero-forcing schemes. 

\subsection{Subnetworks}
It will be useful in each of the achievability and converse proofs provided in this work to treat the network as a set of equal-sized subnetworks; each consisting of $s$ consecutive BS-MT pairs. We use ${\cal L}_k$ to denote the $k^{th}$ subnetwork (set of indices of users in the subnetwork). We define ${\alpha}_{k} = s(k-1)+1$ to denote the first index of each subnetwork ${\cal L}_k$. In this sense, ${\cal L}_k$ is topologically below ${\cal L}_{k-1}$, i.e. mobile terminals from ${\cal L}_k$ are connected to some base stations in ${\cal L}_{k-1}$.

We also make the following definitions.
\begin{defn}
\label{defn:SO}
	We say that the considered transmission scheme relies on \textbf{Subnetwork-only decoding} if words originating in a subnetwork can only be decoded in the same subnetwork. 
\end{defn}
We also define \textbf{Subnetwork-only downlink decoding} and \textbf{Subnetwork-only uplink decoding} to denote that Subnetwork-only decoding is used either for downlink or uplink, respectively.

Finally, we make the following definitions for any zero-forcing message passing decoding scheme in the uplink.
\begin{defn}
For an uplink zero-forcing scheme, if there exist decoding pairs $(\text{MT } i, \text{BS } j)$ such that the mobile terminal $\text{MT }i$ is in ${\cal L}_k$ and the base station $\text{BS }j$ is in ${\cal L}_m, m<k$, then we say that ${\cal L}_k$ \textbf{borrows} base station $j$ from ${\cal L}_m$. 
\end{defn}
We further define $\delta_k$ to denote the number of base stations that ${\cal L}_k$ borrows from ${\cal L}_{k-1}$ to help decode words from ${\cal L}_k$. 
\begin{defn}
For an uplink zero-forcing scheme, if there exist consecutive base stations in ${\cal L}_{k-1}$ indexed by $(\alpha_{k} - \mu_k, \alpha_{k} - \mu_k + 1, \cdots, \alpha_{k} - 1 )$ such that no words can be decoded at these base stations due to the cell association constraint being tightly met in ${\cal L}_{k}$, we say that ${\cal L}_k$ \textbf{blocks} the $\mu_k$ base stations in ${\cal L}_{k-1}$.
\end{defn}

\section{Prior Work: Downlink-Only Scheme}\label{sec:dl}
In~\cite{ElGamal-Annapureddy-Veeravalli-IT14}, the considered setting was studied for only downlink transmission. When restricting our choice of coding scheme to zero-forcing schemes, the puDoF value was characterized as,
\begin{equation}\label{eq:dlzf}
\tau_D^{\textrm{zf}}(L,N_c)=\frac{2N_c}{2N_c+L},
\end{equation}
and the achieving cell association was found to be the following. The network is split into subnetworks; each with consecutive $2N_c+L$ transmitter-receiver pairs. The last $L$ transmitters in each subnetwork are inactive to avoid inter-subnetwork interference. The zero-forcing scheme aims to deliver $2N_c$ messages free of interference in each subnetwork, so that the achieved puDoF value is as in~\eqref{eq:dlzf}. In order to do that with a cooperation constraint that limits each message to be available at $N_c$ transmitters, we create two Multiple Input Single Output (MISO) Broadcast Channels (BC) within each subnetwork; each with $N_c$ transmitter-receiver pairs, and ensure that interference across these channels is eliminated. We now discuss the cell association in the first subnetwork, noting that the remaining subnetworks follow an analogous pattern. The first MISO BC consists of the first $N_c$ transmitter-receiver pairs. For each $i\in\{1,2,\cdots,N_c\}$, message $W_i$ is associated with base stations with indices in the following set, ${\cal C}_i=\{i,i+1,\cdots,N_c\}$. The second MISO BC consists of the $N_c$ transmitters with indices in the set $\{N_c+1,N_c+2,\cdots,2N_c\}$ and the $N_c$ receivers with indices in the set $\{N_c+L+1,N_c+L+2,\cdots,2N_c+L\}$. For each $i\in\{N_c+L+1,N_c+L+2,\cdots,2N_c+L\}$, message $W_i$ is associated with transmitters that have indices in the set ${\cal C}_i=\{i-L,i-L-1,\cdots,N_c+1\}$. Note that the middle $L$ receivers in each subnetwork are deactivated to eliminate interference between the two MISO BCs. It was shown in~\cite{ElGamal-Annapureddy-Veeravalli-IT14} that the puDoF value of~\eqref{eq:dlzf} achieved by this scheme is the best achievable value in the downlink using the considered cooperation constraint and zero-forcing schemes (see also \cite[Chapter $6$]{Veeravalli-ElGamal-Cambridge} for an elaborate illustration).

\section{Main Results}\label{sec:results}
We provide the main results in this section. First, we characterize the average uplink zero-forcing puDoF as follows.

\begin{thm}
\label{thm:uplink}
The zero-forcing asymptotic puDoF for the uplink is characterized as follows: 
\begin{equation}\label{eq:ulzf}
\tau_U^{\textrm{zf}}(L,N_c) =
            \begin{cases}
                1 & L + 1 \leq N_c, \\
                \frac{N_c + 1}{L + 2} & \frac{L}{2} \leq N_c \leq L, \\
                \frac{2N_c}{2N_c + L} & 1 \leq N_c \leq \frac{L}{2} - 1.
            \end{cases}
\end{equation}
\end{thm}
\begin{IEEEproof}
The proof is provided in Section~\ref{sec:ul}.
\end{IEEEproof}
We then consider the problem of characterizing the average zero-forcing puDoF across both uplink and downlink.
\begin{thm}
\label{thm:avguldl}
The following inner bounds are achievable for the average uplink-downlink puDoF using the zero-forcing schemes described in Section \ref{sec:model}:
\begin{equation}\label{eq:uldlzf2}
\tau^{\textrm{zf}}(L,N_c) \geq
            \begin{cases}
                \frac{1}{2}\left(1 + \gamma_{D}(N_c, L)\right) & {L + 1} \leq N_c, \\
                \frac{2N_c}{2N_c + L} & 1 \leq N_c \leq {L},
            \end{cases}
\end{equation}
where \(\gamma_{D}(N_c, L)\) is the downlink component of the puDoF when $N_c \geq L+1$, and is given by

\begin{equation}\label{eq:dlzf2}
\gamma_{D}(L,N_c) =
            \frac{2\left(\left\lceil \frac{L + 1}{2} \right\rceil + N_c - (L + 1)\right)}{L + 2\left(\left\lceil \frac{L + 1}{2} \right\rceil + N_c - (L + 1)\right)}.
\end{equation}
Further, the inner bound in \eqref{eq:uldlzf2} is tight when $N_c \leq \frac{L}{2}$. More precisely,
\begin{equation}\label{eq:uldllby2}
    \tau^{\textrm{zf}}(L,N_c)=\frac{2N_c}{2N_c+L}, \quad \forall N_c \leq \frac{L}{2}.
\end{equation}
\end{thm}
\begin{IEEEproof}
The proof is available in Section~\ref{sec:uldl}.
\end{IEEEproof}
We then characterize the zero-forcing optimal downlink scheme when full coverage associations are used, and note that these associations lead to a unity uplink puDoF.
\begin{thm}
\label{thm:dlunityul}
The optimal zero-forcing downlink puDoF when we have a full coverage association and $N_c > L$ is characterized as, 
\begin{equation}
    \gamma_{D}(L,N_c) =
            \frac{2\kappa}{2\kappa+L},
\end{equation}
where $\kappa = \epsilon + N_c - (L+1)$, and $\epsilon = \left\lceil\frac{L+1}{2}\right\rceil$.
\end{thm}
\begin{IEEEproof}
The proof is available in Section~\ref{sec:fc}.
\end{IEEEproof}
\begin{remark}
One observes that the optimal downlink puDoF of $\frac{2\kappa}{2\kappa + L}$ here has a very similar expression to that of the downlink-only optimal puDoF $\frac{2N_c}{2N_c + L}$. We recall that in the downlink-only analysis, the term $N_c$ was the limit of how many times a word could be associated with a base station during the downlink. When we share associations between uplink and downlink, and use a full coverage association, we essentially reduce that constraint during the downlink from $N_c$ to $\kappa = \left\lceil\frac{L+1}{2}\right\rceil + N_c - (L+1)$.
\end{remark} 
Finally, we demonstrate the information-theoretic optimality of the inner bounds in Theorem~\ref{thm:avguldl} for Wyner's linear networks, i.e., when $L=1$.
\begin{thm}\label{thm:lone}
For Wyner's linear network, the average asymptotic puDoF across both uplink and downlink, is given by,
\begin{equation}\label{eq:lone}
   \tau(L=1,N_c) =
   \begin{cases}
   \frac{1}{2}\left(1 + \frac{2(N_c - 1)}{1 + 2(N_c - 1)}\right) = \frac{4N_c - 3}{4N_c -2}, & N_c \geq 2,\\
   \frac{2}{3}, & N_c=1.
   \end{cases}
\end{equation} 
\end{thm}
\begin{IEEEproof}
The proof is available in Section~\ref{sec:lone}.
\end{IEEEproof}
\section{Uplink-Only Scheme}\label{sec:ul}
We provide the proof of Theorem~\ref{thm:uplink} in the following two subsections.
\subsection{Proof of Achievability}\label{sec:uplink_ach}
The cell association that is used to achieve the puDoF values stated in Theorem~\ref{thm:uplink} is as follows. When \( N_c \geq L+1 \), each mobile terminal is associated with the \(L + 1\) base stations connected to it. The last base station in the network, with index \(K\), decodes the last message and then passes it on to the \(L\) other base stations connected to the \(K^{th}\) mobile terminal, eliminating all interference caused by that mobile terminal. Each preceding base station then decodes its message and passes it on to the other base stations, eliminating the interference caused by the message. Thus, one degree of freedom is achieved for each user.\par

In the second range \(\frac{L}{2} \leq N_c \leq L\), the cell association that is used to achieve a puDoF value of \(\frac{N_c + 1}{L + 2}\) is as follows. The network is split into subnetworks, each with consecutive \(L+2\) transmitter-receiver pairs. In each subnetwork, the last \(N_c + 1\) words are decoded. We now define the cell associations for the first subnetwork. For each \(i \in \{L+2, L+1, \cdots, L + 2 - N_c + 1\}\), message \(W_i\) is associated with base stations \(\{i,i-1,\cdots,L + 2 - N_c + 1\} \subseteq {\cal C}_i \). Thus, the last \(N_c\) words can be decoded while eliminating interference between them. The base stations with indices in the set \(\{2,3,\cdots,L + 2 - N_c\}\) are inactive as there is interference from the last transmitter in the subnetwork which cannot be eliminated. The first base station decodes \(W_{L+2-N_c}\). To eliminate the interference caused by the transmitters with indices in the set \({\cal S} = \{L + 2 - N_c + 1, L + 2 - N_c + 2, \cdots, L + 1\}\) at the first base station of the subnetwork, we add the first base station to each \({\cal C}_i, \forall i \in {\cal S}\). Now for messages with indices in the set \({\cal S}\), we have used \(\beta_i = 2 + i - \left(L + 2 - N_c + 1\right)\) associations up to this point; the factor of two comes from the base station resolving \(W_i\) and the first base station of the subnetwork. But each transmitter with indices in the set \({\cal S} \text{\textbackslash} \{L + 1\} \) also interferes with the subnetwork directly preceding this subnetwork. \(\forall i \in {\cal S} \text{\textbackslash} \{L + 1\}\), the message \(W_i\) interferes with the bottom most \(L + 1 - i\) base stations of the preceding subnetwork, which is precisely the number of associations left for the respective message, i.e. \(N_c - \beta_i = L + 1 - i \), thus inter-subnetwork interference can be eliminated at those base stations. \par

\begin{figure}[t!]
    \centering
    \subfloat[][$N_c < \frac{L}{2}$, $N_c$ = 2, $L$ = 5]{\includegraphics[width=0.5\columnwidth]{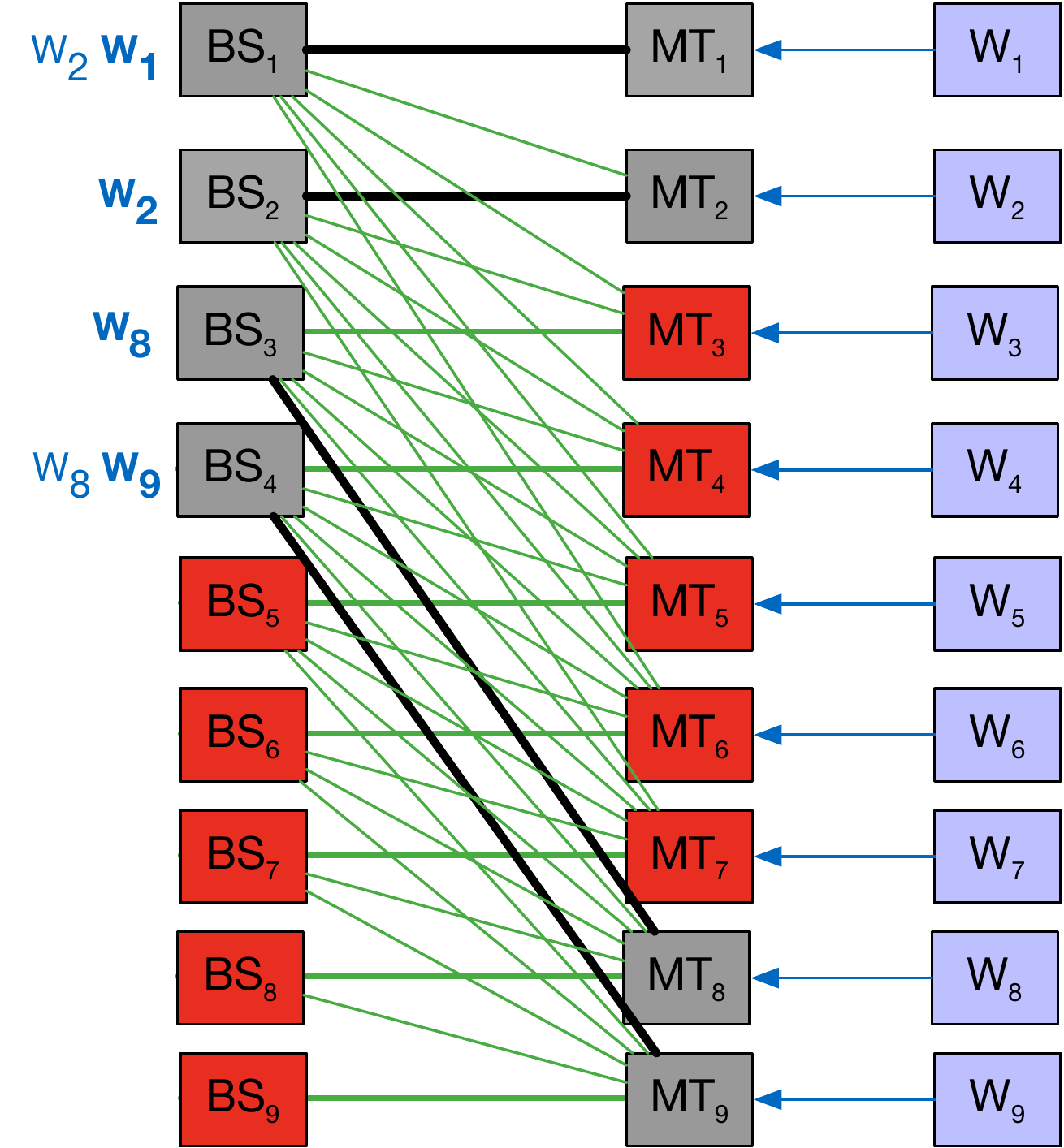}\label{fig:ul_n_c_leq_L_2}}
     \subfloat[][$L+1 \geq N_c \geq \frac{L}{2}$, $N_c$ = 3, $L$ = 3]{\includegraphics[width=0.5\columnwidth]{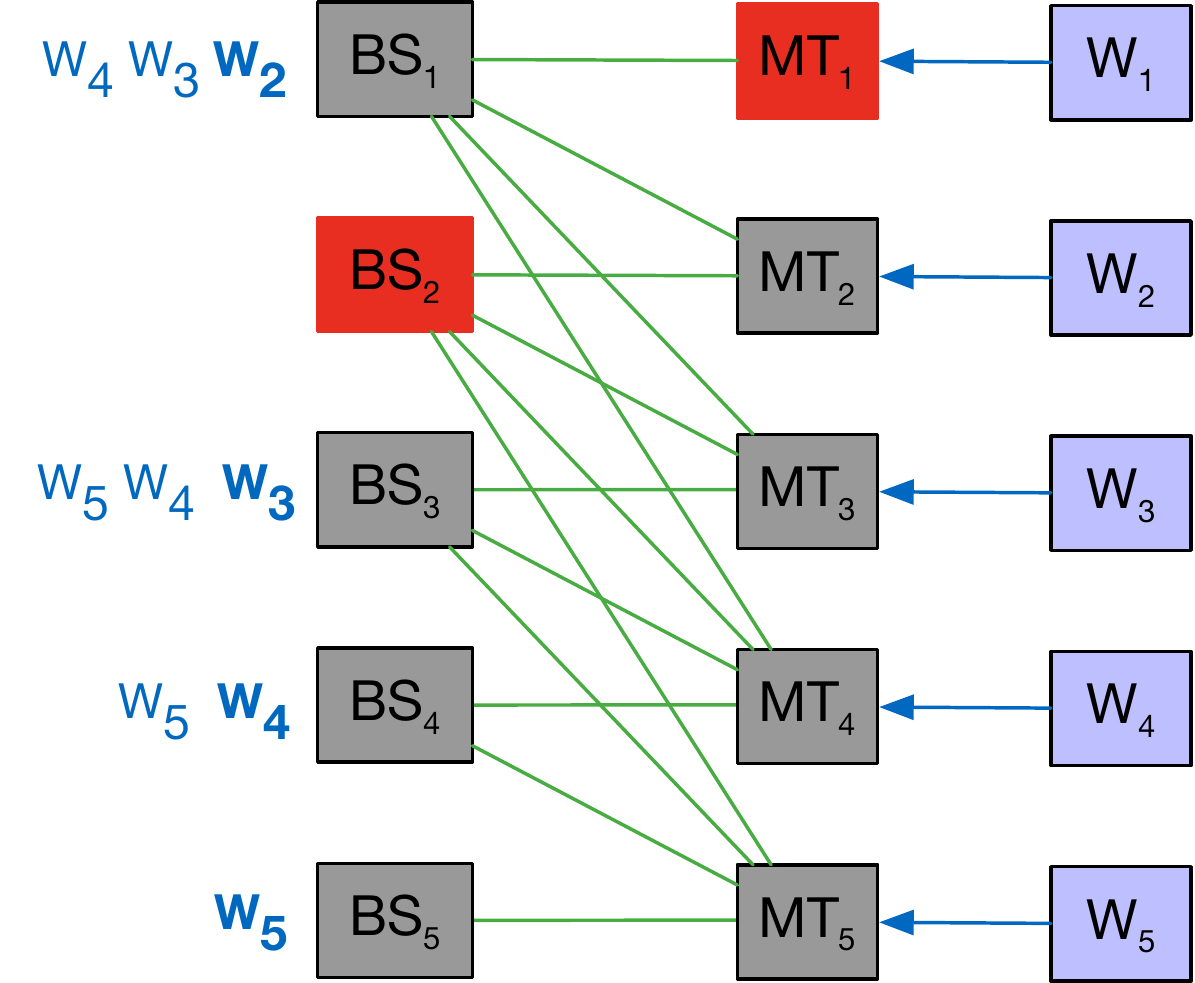}\label{fig:ul_n_c_geq_L_2}}
    \caption{Schemes for Uplink when $N_c \leq L+1$}
    \label{fig:n_c_geq_L}
    \label{fig:msgpass}
\end{figure}

In the third range \(1 \leq N_c \leq \frac{L}{2} - 1\), the cell association that is used to achieve the lower bound of \(\frac{2N_c}{2N_c + L}\) is similar to the one described in Section \ref{sec:dl} for the downlink. The network is split into disjoint subnetworks; each with consecutive \(2N_c + L\) transmitter-receiver pairs. For the uplink, we consider two sets of indices for transmitters \({\cal A}_T = \{1,2,\cdots, N_c\}\) and \({\cal B}_T = \{N_c + L + 1,N_c + L + 2\cdots, 2N_c + L\}\), and corresponding sets of receiver indices \({\cal A}_R = \{1,2,\cdots, N_c\}\) and \({\cal B}_R = \{N_c + 1,N_c + 2, \cdots, 2N_c\}\). For each \(i \in {\cal A}_T\), the message \(W_i\) is associated with the receivers receiving it with indices in \({\cal A}_R\). Receiver \(i\) decodes $W_i$ and the other associations in ${\cal C}_i$ exist for eliminating interference. Similarly, for each \(j \in {\cal B}_T\), the message \(W_j\) is associated with the receivers receiving it with indices in \({\cal B}_R\), but here receiver \(j - L\) decodes $W_j$ and the other associations in ${\cal C}_j$ are for eliminating interference. We illustrate the described schemes in Figure~\ref{fig:msgpass}. 

We observe that if we were not restricted to the zero-forcing coding scheme, then for the third range, we could achieve \(\frac{1}{2}\) puDoF using asymptotic interference alignment~\cite{Cadambe-IA}, which is higher than the value achieved by zero-forcing. The next subsections complete the proof of Theorem \ref{thm:uplink}.

\subsection{Converse Proof}\label{sec:conv}

In this section, we provide a converse proof for the second range of \eqref{eq:ulzf}. More precisely, we show that the following holds.
\begin{equation}\label{eq:ulzf3}
\tau_U^{\textrm{zf}} (L,N_c) = \frac{N_c+1}{L+2}, \qquad \frac{L}{2} \leq N_c \leq L.
\end{equation}
Note that the proof for the remaining case when $N_c < \frac{L}{2}$ follows similarly, and is provided below in Section \ref{sec:convul2}. We start by proving the case when $N_c=L$, The optimal zero-forcing puDoF for the uplink is:

\begin{equation}\label{eq:ulzf2}
\tau_U^{\textrm{zf}}(L,L) =           
                \frac{L + 1}{L + 2}.\\
\end{equation}

We begin by dividing the network into subnetworks of \(L + 2\) consecutive transmitter-receiver pairs. We observe that in any subnetwork, if we have $N_c + 1=L+1$ consecutive active receivers (base stations), then the transmitter connected to all these receivers must be inactive, because a message's interference cannot be canceled at $N_c$ or more receivers. Let \(\Gamma_{BS}\) be the set of subnetworks where all \(N_c + 2\) receivers are active, and \(\Phi_{BS}\) be the set of subnetworks with at most \(N_c\) active receivers. Similarly, let \(\Gamma_{MT}\) and \(\Phi_{MT}\) be the subnetworks with $N_c+2$ active transmitters and at most $N_c$ active transmitters, with respect to order. To be able to achieve a higher puDoF than \eqref{eq:ulzf2}, it must be true that both conditions hold: $| \Gamma_{BS}|>| \Phi_{BS}|$ and $| \Gamma_{MT}|>| \Phi_{MT}|$. Now note that for any subnetwork that belongs to \(\Gamma_{BS}\), at most \(N_c\) transmitters will be active, because the interference caused by any message cannot be canceled at $N_c$ or more receivers. Hence $\Gamma_{BS} \subseteq \Phi_{MT}$. Further, the same logic applies to conclude that for any subnetwork with $N_c+1$ active receivers, the number of active transmitters is at most $N_c+1$, and hence $\Gamma_{MT} \subseteq \Phi_{BS}$. It follows that if $|\Gamma_{BS}|>|\Phi_{BS}|$, then $|\Gamma_{MT}| < |\Phi_{MT}|$, and hence the statement in \eqref{eq:ulzf2} is proved.

%
To prove that $\tau_U^{\textrm{zf}}(L,N_c)=\frac{N_c + 1}{L + 2}$ when $\frac{L}{2} \leq N_c < L$, we use the following lemmas:

\begin{lemma}
\label{exclusitivityclm}
	For any zero-forcing scheme, one of the following is true for any two decoding pairs $(\text{MT } i_1, \text{BS } j_1)$ and $(\text{MT } i_2, \text{BS } j_2)$: $j_2 \notin \{i_1, i_1 -1, \cdots, i_1 - L\}$ or $j_1 \notin \{i_2, i_2 -1, \cdots, i_2 - L\}$.
\end{lemma}
\begin{IEEEproof}
If the claim were not true, i.e. $j_2 \in \{i_1, i_1 -1, \cdots, i_1 - L\}$ and $j_1 \in \{i_2, i_2 -1, \cdots, i_2 - L\}$, then $W_{i_1}$ and $W_{i_2}$ would interfere with one another and could not be decoded using the zero-forcing scheme. This follows from the definition of zero-forcing message passing decoding, first introduced in~\cite{Ntranos-arxiv-CIA}.
\end{IEEEproof}
From Lemma \ref{exclusitivityclm}, we have the following corollary:
\begin{corollary}
\label{orderingclm}
	For any two decoding pairs $(\text{MT }{i_1}, \text{BS }{j_1})$ and $(\text{MT }{i_2}, \text{BS }{j_2})$ in a zero-forcing scheme, if $i_1 > i_2$ then $j_1 > j_2$ and vice versa.
\end{corollary}
We also have the following lemma:
\begin{lemma}
\label{setclm}
	For any set $\cal L \subseteq [K]$ of $L + 1$ consecutive indices, at most $N_c$ messages with indices in ${\cal L}$ can be decoded at base stations with indices in ${\cal L}$ for any zero-forcing scheme.
\end{lemma}
\begin{IEEEproof}
We prove this claim by contradiction. If $N_c + 1$ or more messages with indices in ${\cal L}$ are decoded at base stations with indices in ${\cal L}$, then at least one of the source mobile terminals would be associated with more than $N_c$ base stations. This violates the constraint in~\eqref{eq:backhaul_constraint}.
\end{IEEEproof}
We now present a lemma that forms the foundation for the converse argument that is presented in this section.
\begin{lemma}
\label{borrowingclm}
    If we consider a partitioning of the network users into subnetworks; each of size $L+2$. For the $k^{\textrm{th}}$ subnetwork ${\cal L}_k$, where the largest indexed $y_k \geq 0$ base stations are blocked or borrowed. If ${\cal L}_k$ has $N_c + 1 + (x_k - y_k)$ active mobile terminals, where $x_k > 0$ and $x_k \geq y_k$, then the number of base stations that are blocked or borrowed in the preceding subnetwork ${\cal L}_{k-1}$ is at least $\max\{y_k,1\}$. 
\end{lemma}
\begin{IEEEproof}
We will consider two cases: $y_k = 0$ and $y_k > 0$.
For the case where $y_k = 0$, if ${\cal L}_k$ has $N_c + 1 + x_k$ active mobile terminals, where $x_k > 0$, then by Lemma \ref{setclm}, we have that the largest indexed $L+1$ base stations can only decode a maximum of $N_c$ words. As the size of the considered subnetwork is $L+2$, we have that we can decode a maximum of $N_c + 1$ words in ${\cal L}_k$, thus we will need to borrow base stations to decode the extra $x_k$ words.
For the case where $y_k > 0$, we realize that there are only $L+2 - y_k$ base stations to decode words originating in ${\cal L}_k$. 
If none of the largest indexed $y_k$ mobile terminals were to be active, then the largest indexed active mobile terminal in ${\cal L}_k$ would interfere with the largest indexed $y_k-1$ base stations in ${\cal L}_{k-1}$. Consequently, the second largest indexed active mobile terminal in ${\cal L}_k$ - with index $m_k$ - will interfere with the largest indexed $y_k$ base stations in ${\cal L}_{k-1}$. We then have two cases: If $m_k$ meets its association constraint, then $y_k$ base stations from ${\cal L}_{k-1}$ are blocked or borrowed. If $m_k$ does not meet its association constraint, and there are $N_c-1$ active mobile terminals in ${\cal L}_k$ with a lower index than $m_k$, then one of these mobile terminals will have its message decoded at a base station with an index lower than $\alpha_k-y_k$, and hence, it follows from Lemma~\ref{orderingclm} that $y_k$ base stations in ${\cal L}_{k-1}$ will blocked or borrowed.
It is possible for a subset of the largest indexed $y_k$ mobile terminals to be active; assume without loss in generality that only one of them is active and this mobile terminal is MT $j_k$. If $j_k$ is $\alpha_k + L + 1$, then consider the second largest indexed active mobile terminal MT $m_k$. Now by Lemma \ref{orderingclm}, we have that $m_k$ is at most $\alpha_k + L - y_k$. Hence, MT $m_k$ is connected to at least all the base stations in ${\cal L}_k$ which are not decoding $W_{j_k}$ or are blocked/borrowed by ${\cal L}_{k+1}$. It is also connected to at least $y_k$ base stations in ${\cal L}_{k-1}$. As MT $m_k$ was the second largest indexed active mobile terminal, it will meet its association constraint, thus at least $y_k$ base stations in ${\cal L}_{k-1}$ will be borrowed or blocked. The statement of the lemma then follows.
\end{IEEEproof}

Immediately from Lemma \ref{borrowingclm}, we have that \emph{subnetwork-only uplink decoding}, i.e. when words from a subnetwork are only allowed to be decoded in the same subnetwork, can not decode more than $N_c + 1$ words in a subnetwork of size $L+2$.
Our proof will be based on the concept that to exceed the inner bound described in \eqref{eq:ulzf}, at least one subnetwork of $L+2$ consecutive MT-BS pairs must have more than $N_c + 1$ active mobile terminals. Say this subnetwork is ${\cal L}_k$. And any such subnetwork must \emph{borrow} base stations from the subnetwork above it to decode words corresponding to its own mobile terminals. We define a \emph{best case} scenario for inter-subnetwork interference. A best case scenario is where the interference from one subnetwork's (e.g., ${\cal L}_k$) mobile terminals to another subnetwork's (e.g., ${\cal L}_{k-1}$) base stations is focused on the bottom most base stations. This is defined as the best case scenario because from Lemma \ref{orderingclm}, we know that for ${\cal L}_{k-1}$'s own mobile terminals to be decoded in ${\cal L}_{k-1}$, we need base stations that are indexed outside the range of the interference from the active mobile terminals of ${\cal L}_k$. 

We want to show that $\tau_U^{\textrm{zf}}(L,N_c) \leq \frac{N_c + 1}{L + 2}$ when $L > N_c \geq \frac{L}{2}$. It follows from the pigeonhole principle that to break this bound, there must be at least one subnetwork (say  ${\cal L}_k$) where we have $N_c + 1 + x_k$ active mobile terminals, $x_k > 0$. Now by Lemma \ref{setclm}, we have that ${\cal L}_{k}$ must borrow at least $x_k$ base stations from ${\cal L}_{k-1}$, thus $x_k \leq \delta_k$. We now consider possible cases for the value of $\delta_k$. Note that it follows from the network topology and the defined cell association constraint that $\delta_k \leq N_c$.

If $\delta_k = 1$, thus $x_k = 1$, so  ${\cal L}_k$ has $N_c + 2$ active mobile terminals. As ${\cal L}_k$  is borrowing one base station, say base station $j$, $N_c + 1$ words must have been decoded in  ${\cal L}_k$. By Lemma \ref{setclm}, there exists at least one decoding pair $(\text{MT }i,\text{BS }n)$ where $i,n \geq \alpha_k$, such that $\text{BS }n$ is not connected to the highest indexed active mobile terminal in ${\cal L}_k$. Due to the size of the subnetwork, this forces $n = \alpha_k$. Hence, mobile terminal $i$'s word is decoded at the first base station of ${\cal L}_k$. By Lemma \ref{exclusitivityclm}, this implies that $j \notin \{i, i-1, ..., i-L\}$. It follows that $i \leq \alpha_{k} + (L + 2 - (N_c + 1))$, making $j \leq \alpha_k - N_c = \alpha_{k-1} + L+2-N_c$. Let the number of base stations left in ${\cal L}_{k-1}$ that can decode words originating in ${\cal L}_{k-1}$ be $\theta$. It follows that $\theta \leq L + 2 - N_c $. Thus we have that $L+2 - \theta \geq N_c$ base stations are blocked or borrowed in ${\cal L}_{k-1}$.

Now by Lemma \ref{borrowingclm}, we have that for ${\cal L}_{k-1}$ to have at least $N_c+1$ active mobile terminals, it would have to borrow or block base stations from ${\cal L}_{k-2}$. We do not consider the case where ${\cal L}_{k-1}$ has less than $N_c$ active mobile terminals as that would force the average number of active mobile terminals across ${\cal L}_k$ and ${\cal L}_{k-1}$ to be less than or equal to $N_c$, and we could just restart our argument from ${\cal L}_{k-2}$. By Lemma \ref{borrowingclm}, we have that ${\cal L}_{k-1}$ will block or borrow at least $L + 2 - \theta$ base stations in ${\cal L}_{k-2}$. So now if $\delta_{k-1} = 1$ we have that the number of base stations blocked or borrowed in ${\cal L}_{k-1}$ is the same as the number of base stations blocked or borrowed in ${\cal L}_{k-2}$, and thus this borrowing/blocking will continue till either we stop borrowing at some some subnetwork ${\cal L}_i$ or we reach ${\cal L}_1$. In the former case, we will get the case that the overall average between ${\cal L}_k$ and ${\cal L}_i$ is $N_c + 1$, as in ${\cal L}_i$ we will have at most $N_c$ active mobile terminals. In the latter case, we have that we only get one extra active mobile terminal over the whole network, which will not affect the asymptotic per user DoF. If $\delta_{k-1} > 1$, we have a similar argument to the one where $\delta_k > 1$, which is discussed next.

When $\delta_k > 1$, we have a similar argument as described in the previous paragraph. By Lemma \ref{exclusitivityclm}, we have that the largest indexed borrowed base station in  ${\cal L}_{k-1}$ will have to be connected to the smallest indexed active mobile terminal of  ${\cal L}_k$, and it is not connected to any other active mobile terminal in ${\cal L}_k$. As the index of the smallest indexed active mobile terminal in ${\cal L}_k$ is at most $\alpha_{k} + (L + 2 - (N_c + 1 + x_k)) - 1$, we have that the index of the lowest borrowed base station in ${\cal L}_{k-1}$ is at most $\alpha_{k-1} + (L+2 - N_c - x_k)$. Therefore, the number of base stations in ${\cal L}_{k-1}$ that can decode words originating in ${\cal L}_{k-1}$ can be at most $L + 3 - N_c - x_k$. These available base stations must at least decode $N_c + 1 + (1 - x_k)$ mobile terminals' words to have an average greater than $N_c + 1$ active mobile terminals per subnetwork over ${\cal L}_k$ and ${\cal L}_{k-1}$ without ${\cal L}_{k-1}$ borrowing base stations from ${\cal L}_{k-2}$. This cannot happen when $L + 2 - N_c - x_k < N_c + 1 + 1 - x_k$, which corresponds to $N_c \geq \frac{L+1}{2}$. Hence, the condition $N_c \geq \frac{L+1}{2}$ implies that  ${\cal L}_{k-1}$ has to borrow at least one base station from  ${\cal L}_{k-2}$, which presents an iterative argument akin to the one shown when $\delta_k=1$.

In order to achieve a case where ${\cal L}_{k-1}$ does not have to borrow base stations from ${\cal L}_{k-2}$, our best case scenario guides us to find the first mobile terminal in ${\cal L}_{k-1}$, which is connected to at most  $x_k - 2$ base stations that are being borrowed by ${\cal L}_{k}$, but still connected to at least $N_c + 2 - x_k$ available base stations in ${\cal L}_{k-1}$. Assume that the index of that mobile terminal is $\alpha_{k-1} + \nu$. Clearly, $\nu \leq (L + 2 - N_c - x_k) + (x_k - 2) = L - N_c$. Hence, in ${\cal L}_{k-1}$ we have $N_c + 2 - x_k$ active mobile terminals without borrowing from ${\cal L}_{k-2}$, but mobile terminal ${\alpha_{k-1} + \nu}$ has already used up all its associations and it is connected to at least $N_c$ base stations in ${\cal L}_{k-2}$. Hence, ${\cal L}_{k-2}$ has a maximum of $L + 2 - N_ c \leq N_c + 2$ base stations available to decode more words, and we need at least $N_c + 1$ words to be decoded here, which can be done, but this would imply that at least two mobile terminals are associated with $N_c$ base stations. These two mobile terminals are indexed higher than $\kappa$, where $\kappa = \alpha_{k-2} + L + 1 - (N_c + 1)$. Hence, ${\cal L}_{k-2}$ blocks at least $N_c$ of the bottom most $L$ base stations in ${\cal L}_{k-3}$, and one can see that each further subnetwork blocks at least one base station from the preceding subnetwork for the average number of active mobile terminals per subnetwork to remain above $N_c+1$. If say ${\cal L}_i$ does not block any base stations in ${\cal L}_{i-1}$, then ${\cal L}_i$ can have at most $N_c$ active mobile terminals decoded in ${\cal L}_i$. It follows that either ${\cal L}_i$ borrows from ${\cal L}_{i-1}$ or only has $N_c$ active mobile terminals. If ${\cal L}_i$ borrows from ${\cal L}_{i-1}$, we have a similar iterative argument as shown above. Otherwise, ${\cal L}_i$ has only $N_c$ active mobile terminals, making the average number of active mobile terminals through the considered $k-i$ subnetworks ${N_c + 1}$ per subnetwork. Hence, each subnetwork is blocking base stations in the preceding subnetwork and the number of extra active mobile terminals in the whole network does not scale, and is fixed by the constant $x_k$, which shows that the average number of active mobile terminals asymptotically approaches ${N_c + 1}$ for every subnetwork of size ${L + 2}$.

We have shown that if any subnetwork has more than $N_c + 1$ active mobile terminals when $L \geq N_c \geq \frac{L}{2}$, either the number of extra active mobile terminals does not scale with size of the network, or the average over the whole network remains bounded by $N_c + 1$ active mobile terminals per subnetwork. This forces the asymptotic average number of decoded words per subnetwork to be at most $N_c + 1$, implying that the asymptotic puDoF during the uplink using zero forcing, $\tau_U^{\textrm{zf}}(L,N_c) \leq \frac{N_c + 1}{L + 2}$. We have shown in Section~\ref{sec:uplink_ach} that $\tau_U^{\textrm{zf}}(L,N_c) \geq \frac{N_c + 1}{L + 2}$, implying that $\tau_U^{\textrm{zf}}(L,N_c) = \frac{N_c + 1}{L + 2}$ whenever $\frac{L}{2} \leq N_c \leq L$. 

\subsection{Converse Proof when \( N_c < \frac{L}{2}\)}\label{sec:convul2}

In this section, we provide a converse proof for the third range of \eqref{eq:ulzf}. More precisely, we show that the following holds.
\begin{equation}\label{eq:ulzf4}
\tau_U^{\textrm{zf}} (L,N_c) = \frac{2N_c}{2N_c + L}, \qquad N_c < \frac{L}{2}.
\end{equation}

Similar to Section \ref{sec:conv}, our proof will be based on the concept that to exceed the inner bound described in \eqref{eq:ulzf}, at least one subnetwork of $2N_c + L$ consecutive MT-BS pairs must have more than $2N_c$ active mobile terminals. And any such subnetwork must either \emph{borrow} or \emph{block} base stations from the subnetwork above it to decode words corresponding to its own mobile terminals. 

Consider the first subnetwork which has more than $2N_c$ active mobile terminals. Say this subnetwork is ${\cal L}_k$. 
We first consider the case when $\delta_k = 0$, i.e ${\cal L}_k$ does not borrow any base stations from ${\cal L}_{k-1}$. To help in writing the proof, we first mark three special mobile terminals of ${\cal L}_k$, the largest indexed active mobile terminal MT $\alpha_k + \alpha$, the $(N_c + 1)^{st}$ largest indexed active mobile terminal MT $\alpha_k + \beta$, and the $(N_c + 2)^{nd}$ largest indexed active mobile terminal MT $\alpha_k + \gamma$. By definition, we note that $\alpha - \beta \geq N_c$, and $\beta - \gamma \geq 1$.

It follows from Lemma \ref{setclm} that we have two possible scenarios, the first one being that the $N_c$ largest indexed words are decoded in the $L+1$ base stations connected to MT $\alpha_k + \alpha$, and the second being that for the subnetwork consisting of the $L+1$ indices of base stations connected to MT $\alpha_k + \alpha$, only $x < N_c$ words are decoded.

In the first scenario, we are left with at most $2N_c - 1$ base stations to decode at least $N_c  + 1$ words. Consider MT $\alpha_k + \beta$. If $\beta < \alpha-(L+1)$, then the upper $2N_c-1$ base stations in ${\cal L}_k$ have to decode more than $N_c$ words originating at the upper $2N_c-1$ mobile terminals in ${\cal L}_k$, which is impossible because of Lemma \ref{setclm}. We hence restrict our attention to the case when $\beta \geq \alpha-(L+1)$. We know that either $W_{\alpha_k + \beta}$ is decoded in one of the base stations indexed in the set $\{\alpha_k + \beta - L + 1, \cdots, \alpha_k + \alpha - (L+1)\}$ because of Lemma \ref{setclm}, and by Lemma \ref{orderingclm} no other word indexed in the set $\{\alpha_k + \alpha - (L+1), \cdots, \alpha_k + \beta - 1\}$ is decoded in the set of base stations indexed in $\{\alpha_k + \beta - L, \cdots, \alpha_k + \alpha - (L+1)\}$, or that $W_{\alpha_k + \beta}$ is decoded at BS $\alpha_k + \beta - L$, which would leave at most $N_c - 1$ base stations to decode at least $N_c$ words, which is clearly impossible. 

So we have that $W_{\alpha_k + \beta}$ is decoded at one of the base stations indexed in $\{\alpha_k + \beta - L + 1, \cdots, \alpha_k + \alpha - (L+1)\}$ and no mobile terminal indexed in the set $\{\alpha_k + \alpha - (L+1), \cdots, \alpha_k + \beta - 1\}$ is decoded at the base stations indexed in the set $\{\alpha_k + \beta - L, \cdots, \alpha_k + \alpha - (L+1)\}$. Let the base station that decodes $W_{\alpha_k + \beta}$ be BS $\alpha_k + \eta$. We have two cases for the index $\gamma$: Either $\gamma \leq \eta - 1$ or $\gamma \in \{\eta, \cdots, \beta - 1\}$. If the latter holds, then we observe that Lemma \ref{orderingclm} would imply that $W_{\alpha_k + \gamma}$ is decoded at a BS with an index that is at most $\alpha_k + \beta - L - 1$, but this would leave at most $N_c - 2$ base stations to decode at least $N_c - 1$ words, which is clearly impossible. If the former holds, i.e., MT $\alpha_k + \gamma$ is not below MT $\alpha_k + \eta - 1$, as $\alpha_k + \eta \leq \alpha_k + \alpha - L - 1$, then this leaves at most $2N_c - 2 < L+1$ base stations to decode at least $N_c$ words. Thus, it follows from Lemma \ref{setclm} that at most $2N_c+1$ mobile terminals are active in ${\cal L}_k$, and if that is the case, then ${\cal L}_k$ blocks base stations in ${\cal L}_{k-1}$, specifically it blocks $L+1 - (2N_c - 2) = L - 2N_c + 3$ base stations. The end result for this scenario is depicted in Figure \ref{fig:Lk for 5,2} for $L = 5, N_c = 2$.
\begin{figure}[H]
    \centering
    \includegraphics[height=6cm]{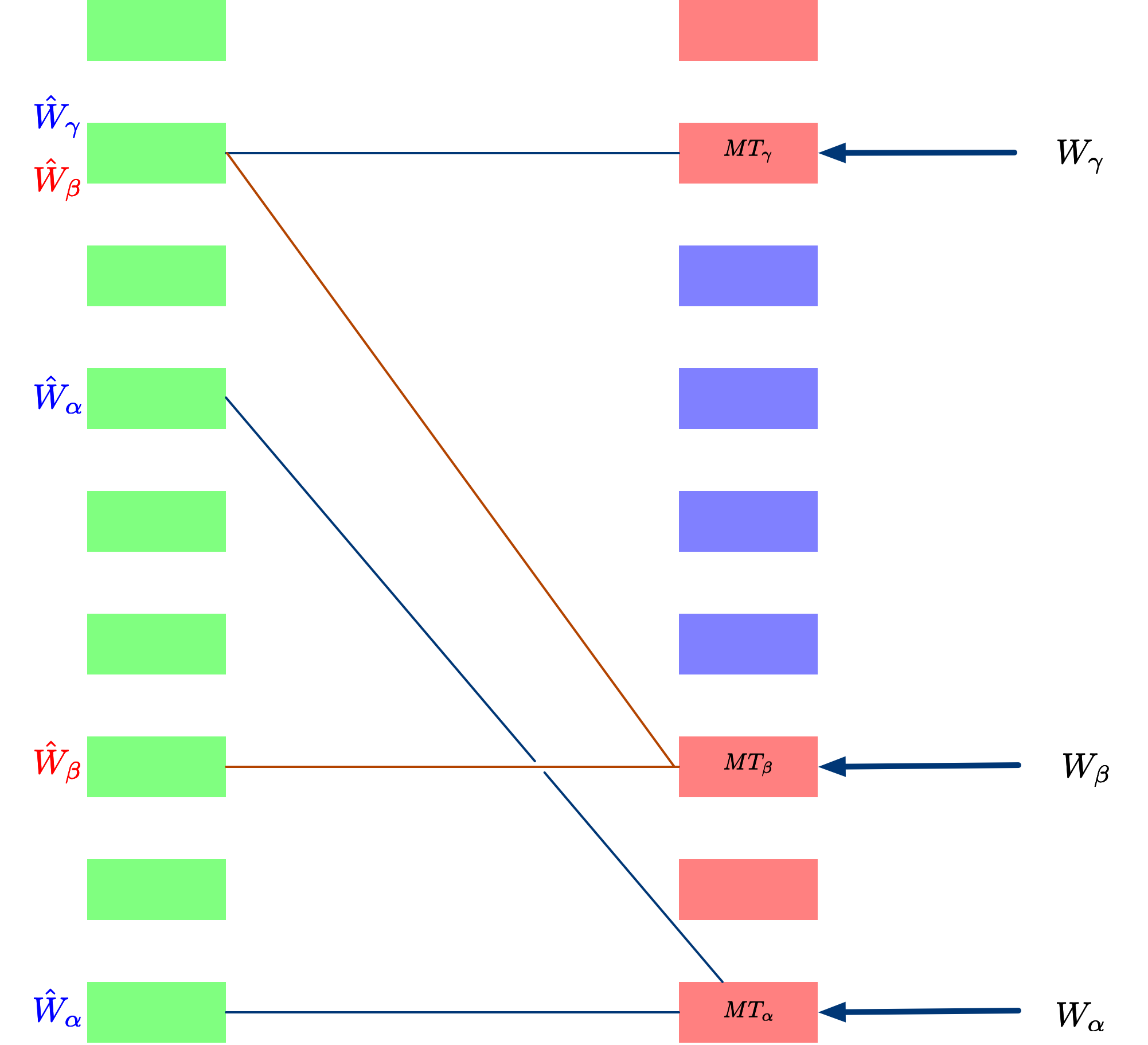}
    \caption{${\cal L}_k$ for $L = 5, N_c = 2$, where the red terminals are the active ones, and the blue terminals are the inactive ones}
    \label{fig:Lk for 5,2}
\end{figure}

In the second scenario, we mark another special mobile terminal, specifically the $N_c^{th}$ largest indexed active mobile terminal, call this MT $\alpha_k + \psi$. So by definition $W_{\alpha_k + \psi}$ is decoded in a base station indexed at most $\alpha_k + \alpha - (L+1)$. Thus, we have that either $W_{\alpha_k + \beta}$ is decoded in one of the base stations indexed in $\{\alpha_k + \beta - L + 1, \cdots, \alpha_k + \alpha - (L+1) - 1\}$ and no other mobile terminal indexed in the set $\{\alpha_k + \alpha - (L+1) - 1, \cdots, \alpha_k + \beta - 1\}$ is decoded in the base stations indexed in $\{\alpha_k + \beta - L, \cdots, \alpha_k + \alpha - (L+1) - 1\}$, or that $W_{\alpha_k + \beta}$ is decoded at base station BS $\alpha_k + \beta - L$, which would leave $N_c - 1$ base stations to decode at least $N_c$ words, which is clearly impossible. So, the former case holds. This implies that MT $\alpha_k + \gamma$ is not below MT $\alpha_k + \alpha - (L+1) - 2$, but this leaves at most $2N_c - 3 < L+1$ base stations to decode the $N_c$ words left to decode. Thus, ${\cal L}_k$ blocks base stations in ${\cal L}_{k-1}$, specifically more than the previous scenario, so moving on we only consider the previous scenario.

We then consider the case where ${\cal L}_k$ blocks $L - 2N_c + 3$ base stations in ${\cal L}_{k-1}$. In addition, ${\cal L}_{k-1}$ needs to decode at least $2N_c$ words to surpass the average uplink puDoF of $\frac{2N_c}{2N_c + L}$, as ${\cal L}_k$ decodes $2N_c + 1$ words. We then have $2N_c + L - (L - 2N_c + 3) = 4N_c -3$ base stations to decode the $2N_c$ words. Similar to the argument we made above for ${\cal L}_k$, we first mark three special mobile terminals, the largest indexed active mobile terminal MT $\alpha_{k-1} + \alpha'$, the $(N_c + 1)^{st}$ largest indexed active mobile terminal MT $\alpha_{k-1} + \beta'$, and the $(N_c + 2)^{nd}$ largest indexed active mobile terminal MT $\alpha_{k-1} + \gamma'$. Now we know that the largest indexed $L - 2N_c + 3$ base stations cannot be used to decode any words, thus $W_{\alpha_{k-1} + \alpha'}$ must either be decoded at a base station indexed in $\{\alpha_{k-1} + \alpha' - (L) + 1, \cdots, \alpha_{k-1} + 4N_c - 3 - 1\}$, or be decoded at the base station BS $\alpha_{k-1} + \alpha' - L$. 

If it were the former and say $W_{\alpha_{k-1} + \alpha'}$ was decoded at BS $\alpha_{k-1} + \delta'$, Lemma \ref{orderingclm} would then force that no mobile terminal in ${\cal L}_{k-1}$ indexed in the set $\{\alpha_{k-1} + \delta', \cdots,\alpha_{k-1} + \alpha' - 1\}$ can be decoded in the interference set of MT $\alpha_{k-1} + \alpha'$. Hence, the second largest active mobile terminal (MT $\alpha_{k-1} + \epsilon'$) would be at most indexed at $\alpha_{k-1} + 4N_c - 5$. Now as $\alpha_{k-1} + \alpha' - (L+1) \in \{\alpha_{k-1} + \epsilon' - L, \cdots, \alpha_{k-1} + \epsilon'\}$, we observe that it is either the case that MT ${\alpha_{k-1}+\epsilon'}$ is associated with the maximum of $N_c$ base stations, or it is the case that at most $N_c - 1$ words are decoded at base stations in the interference set of MT $\alpha_{k-1} + \epsilon'$. Thus, we have at most $4N_c - 4 - (L+1) < 2N_c - 5 < L+1$ base stations to decode at least $N_c - 1$ words. Now by Lemma \ref{orderingclm}, we have that $W_{\alpha_{k-1} + \gamma'}$ must be decoded at a base station indexed at most $\alpha_{k-1} + 2N_c - 7$, thus we observe that the interference set of MT $\alpha_{k-1} + \gamma'$ would allow at most one word to be decoded in the largest indexed $L - 2N_c + 7$ base stations in ${\cal L}_{k-2}$.

If it were the latter, we would observe that we would be left with exactly $2N_c-1$ base stations to decode $2N_c - 1$ words. In addition due to Lemma 1, $\alpha_{k-1} + \beta'$ would be exactly $\alpha_{k-1} + N_c - 1$. This would force that MT $\alpha_{k-1} + \beta'$ blocks exactly $L + 1 - N_c$ base stations in ${\cal L}_{k-2}$. As $N_c < \frac{L}{2}$, we have that $L + 1 - N_c \geq 3$, thus we block at least three base stations, which is at least the same number of base stations that ${\cal L}_{k}$ blocked in ${\cal L}_{k-1}$, thus we can just reuse our argument for ${\cal L}_{k-1}$ when we consider ${\cal L}_{k-2}$. 

\begin{figure}[H]
    \centering
    \includegraphics[height=8cm]{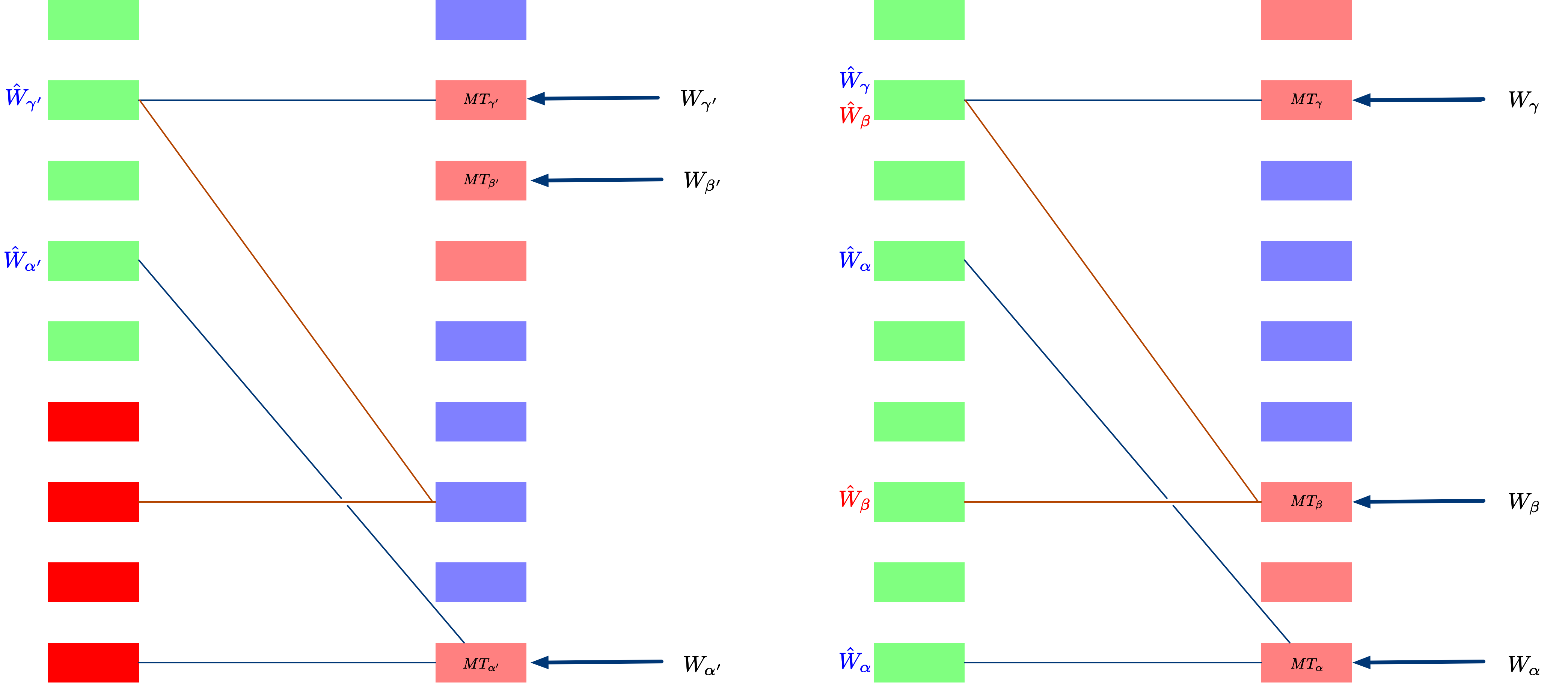}
    \caption{${\cal L}_k$ and ${\cal L}_{k-1}$ for $L = 5, N_c = 2$, where the red base stations are the blocked base stations.}
    \label{fig:chain_interference}
\end{figure}

In ${\cal L}_{k-2}$, we again mark three special mobile terminals: MT $\alpha_{k-2} + \alpha''$ for the largest indexed active mobile terminal, MT $\alpha_{k-2} + \beta''$ for the third largest indexed active mobile terminal, and MT $\alpha_{k-2} + \gamma''$ for the $(N_c + 3)^{rd}$ largest indexed active mobile terminal. Now we have two cases for $W_{\alpha_{k-2} + \alpha''}$, it can either be decoded in the set of base stations blocked by ${\cal L}_{k-1}$, or outside this set. In the latter case, we observe that the second largest indexed active mobile terminal, say MT $\alpha_{k-2} + \delta''$, would be decoded at a base station indexed at most MT $\alpha_{k-2} + 4N_c - 7 - 1$. By Lemma \ref{orderingclm}, we would observe that this would force the index of the third largest indexed active mobile terminal to be at most $\alpha_{k-2} + 4N_c - 7 - 2$, thus we are left with at most $4N_c - 9 - (L+1) < 2N_c - 10$ base stations to decode at least $N_c - 3$ words. We observe that compared to the above considered case for ${\cal L}_{k-1}$, the number of available base stations have reduced by at least five, and the number of words left to decode at these base stations have decreased by at most two. This inter-subnetwork interference propagation pattern continues till the first subnetwork, and we observe that the extra mobile terminal we decoded in ${\cal L}_k$ does not add to the average asymptotic puDoF. An example for this inter-subnetwork interference propagation pattern is shown in Figure \ref{fig:chain_interference}.

In Figure \ref{fig:chain_interference}, we observe that the pattern of interference propagation makes it impossible to increase the number of extra active mobile terminals gained in ${\cal L}_k$, as early as ${\cal L}_{k-1}$. It is clearly impossible for ${\cal L}_{k-1}$ to even decode $2N_c$, as MT $\gamma'$  cannot be decoded in ${\cal L}_{k-1}$, thus the overall average DoF for the subnetworks ${\cal L}_k$ and ${\cal L}_{k-1}$ is less than or equal to $\frac{2N_c}{2N_c + L}$, which is exactly the upper bound we proposed.

In the former case we observe that we would be left with at most $4N_c - 9 - (L+1) <2N_c - 10$ base stations to decode $N_c - 2$ words, which leaves us in a more constrained situation than the latter case.

For the case where $\delta_k > 0$, we observe that even if we borrow base stations to decode words originating from ${\cal L}_k$, the largest indexed $L - 2N_c + 3$ base stations in ${\cal L}_{k-1}$ will be either blocked or will not be able to decode a single word from ${\cal L}_{k-1}$ as a result of Lemma \ref{orderingclm}. Thus we are left with at least the same constrained situation as in previous cases. We illustrate the argument with more details in what follows.
 
We follow the same notation that we followed earlier for the case where $\delta_k = 0$. It follows from Lemma \ref{setclm} that we have two possible scenarios for the subnetwork ${\cal L}_k$: The first one being that the $N_c$ largest indexed words are decoded in the $L+1$ base stations connected to MT $\alpha_k + \alpha$, and the second being that for the subnetwork consisting of the $L+1$ indices of base stations connected to MT $\alpha_k + \alpha$, only $x < N_c$ words are decoded.

In the first scenario, consider MT $\alpha_k + \beta$, we know from Lemma 2 that either $W_{\alpha_k + \beta}$ is decoded in one of the base stations indexed in $\{\alpha_k + \beta - L + 1, \cdots, \alpha_k + \alpha - (L+1)\}$, and by Lemma 1 no other word indexed in the set $\{\alpha_k + \alpha - (L+1), \cdots, \alpha_k + \beta - 1\}$ is decoded in the set of base stations indexed in $\{\alpha_k + \beta - L, \cdots, \alpha_k + \alpha - (L+1)\}$, or that $W_{\alpha_k + \beta}$ is decoded at base station BS $\alpha_k + \beta - L$. We first consider the latter case, which would leave only $N_c - 1$ base stations in ${\cal L}_{k}$ to decode any words indexed lower than $\alpha_k + \beta$ originating in ${\cal L}_{k}$. The number of active mobile terminals indexed lower than $\alpha_k + \beta$ is at least $N_c - 1$. Thus, the maximum number of words that ${\cal L}_k$ can decode is $2N_c$, due to the tight constraint that we have exactly $N_c - 1$ base stations left to decode $N_c - 1$ words. This implies that in ${\cal L}_{k-1}$, the largest indexed $L - N_c + 2$ base stations decode at most one word, and this word has to be from ${\cal L}_k$, as ${\cal L}_k$ has at least $2N_c + 1$ active mobile terminals. Thus in ${\cal L}_{k-1}$, at least $L - 2N_c + 1$ of the largest indexed mobile terminals cannot decode any words originating in ${\cal L}_{k-1}$, which is strictly greater than $L - 2N_c + 3$, which is the number of base stations blocked when $\delta_k = 0$, thus we are left in a situation where we have less available base stations in ${\cal L}_{k-1}$ to decode the same number of words. Even if we decode $2N_c - x$ words in ${\cal L}_{k}$ where $x < N_c - 1$, it would still force that in the largest indexed $L - N_c + 2$ base stations, we could only decode $x+1$ words and they all have to originate from ${\cal L}_{k}$. Thus this scenario gives us no more extra decoded words compared to the previous case where $\delta_k = 0$.

Now we consider the case where $W_{\alpha_k + \beta}$ is decoded at one of the base stations indexed in $\{\alpha_k + \beta - L + 1, \cdots, \alpha_k + \alpha - (L+1)\}$ and no mobile terminal indexed in the set $\{\alpha_k + \alpha - (L+1), \cdots, \alpha_k + \beta - 1\}$ is decoded at the base stations indexed in the set $\{\alpha_k + \beta - L, \cdots, \alpha_k + \alpha - (L+1)\}$. For ease in writing, we say that the base station that decodes $W_{\alpha_k + \beta}$ is BS $\alpha_k + \eta$. This implies that we have two cases for MT $\alpha_k + \gamma$: Either its index is at most $\alpha_k + \eta - 1$, or MT $\alpha_k + \gamma$ is indexed in $\{\alpha_k + \eta, \cdots, \alpha_k + \beta - 1\}$. If it is the latter, we observe that Lemma 1 would imply that $W_{\alpha_k + \gamma}$ is decoded at a BS indexed at most $\alpha_k + \beta - L - 1$, but this would leave at most $N_c - 2$ base stations to decode at least $N_c - 1$ words, which is clearly impossible. If it is the former, i.e. $\alpha_k + \gamma$ is at most $\alpha_k + \eta - 1$, as $\alpha_k + \eta$ is at most $\alpha_k + \alpha - L - 1$, then this leaves at most $2N_c - 2 < L+1$ base stations to decode $N_c$ words. But we realize that to decode at least $2N_c+1$ words originating in ${\cal L}_k$, either the largest indexed $L - 2N_c + 3$ base stations in ${\cal L}_{k-1}$ only decode words originating in ${\cal L}_k$, or do not decode any words at all. This leaves us with at least the same constraints we had when $\delta _k = 0$, when the largest indexed $L - 2N_c + 3$ base stations in ${\cal L}_{k-1}$ were blocked, or could not decode any words originating from ${\cal L}_{k-1}$.

In the second scenario, we mark another special mobile terminal, specifically the $N_c^{th}$ largest indexed active mobile terminal, and call this MT $\alpha_k + \psi$. So by definition, $W_{\alpha_k + \psi}$ is decoded in a base station indexed at most $\alpha_k + \alpha - (L+1)$. Thus, we have that either $W_{\alpha_k + \beta}$ is decoded at one of the base stations indexed in $\{\alpha_k + \beta - L + 1, \cdots, \alpha_k + \alpha - (L+1) - 1\}$ and no other mobile terminal indexed in the set $\{\alpha_k + \alpha - (L+1) - 1, \cdots, \alpha_k + \beta - 1\}$ is decoded at the base stations indexed in $\{\alpha_k + \beta - L, \cdots, \alpha_k + \alpha - (L+1) - 1\}$, or that $W_{\alpha_k + \beta}$ is decoded at base station BS $\alpha_k + \beta - L$, which would leave $N_c - 1$ base stations to decode at least $N_c$ words, which clearly would lead to borrowing base stations from ${\cal L}_{k-1}$. Similar to the previous scenario, we have the case where ${\cal L}_k$ borrows base stations to decode words but blocks $L - N_c + 1$ base stations in ${\cal L}_{k-1}$, which is strictly greater than $L - 2N_c + 3$, which was the case when $\delta_k = 0$.
Hence, it remains to consider the case where $W_{\alpha_k + \beta}$ is decoded in one of the base stations indexed in $\{\alpha_k + \beta - L + 1, \cdots, \alpha_k + \alpha - (L+1) - 1\}$ and no other mobile terminal indexed in the set $\{\alpha_k + \alpha - (L+1) - 1, \cdots, \alpha_k + \beta - 1\}$ is decoded in the base stations indexed in $\{\alpha_k + \beta - L, \cdots, \alpha_k + \alpha - (L+1) - 1\}$. This implies that MT $\alpha_k + \gamma$ is at most MT $\alpha_k + \alpha - (L+1) - 2$, but this leaves at most $2N_c - 3 < L+1$ base stations to decode the $N_c$ words left to decode. Thus, ${\cal L}_k$ blocks and/or borrows base stations in ${\cal L}_{k-1}$, specifically more than the previous scenario.

To summarize, even when we allow subnetworks to borrow base stations from neighboring subnetworks, we realize that the number of borrowed base stations still do not allow us to scale the increased number of active words per subnetwork beyond $2N_c$. Thus, the upper bound for the puDoF still remains $\frac{2N_c}{2N_c + L}$.
\section{Average Uplink-Downlink Degrees of Freedom}\label{sec:uldl}
In this section, we present zero-forcing schemes, with the goal of optimizing the average rate across both uplink and downlink for arbitrary values of the connectivity parameter $L$. We first prove the inner bounds provided in Theorem~\ref{thm:avguldl}, and note that
if the inner bound in \eqref{eq:uldlzf2} holds, then the proof of \eqref{eq:uldllby2} follows from Theorem~\ref{thm:uplink} and the result in~\cite{ElGamal-Annapureddy-Veeravalli-IT14} described in Section~\ref{sec:dl}. Hence, we only provide the achievability proof to complete the proof of Theorem~\ref{thm:avguldl}.

The coding scheme that achieves the inner bound for the second range of~\eqref{eq:uldlzf2} is essentially the union of the scheme described in Section \ref{sec:dl} and the scheme that achieves the third range of \eqref{eq:ulzf}. The network is split into disjoint subnetworks; each with consecutive \(2N_c + L\) transmitter-receiver pairs. We consider two sets of base station indices: \({\cal A}_{BS} = \{1,2,\cdots, N_c\}\) and \({\cal B}_{BS} = \{N_c + 1,N_c + 2\cdots, 2N_c\}\), and two sets of mobile terminal indices: \({\cal A}_{MT} = \{1,2,\cdots, N_c\}\) and \({\cal B}_{MT} = \{N_c + L + 1,N_c + L + 2\cdots, 2N_c + L\}\). Now for each \(i \in {\cal A}_{MT}\), \({\cal C}_i={\cal A}_{BS}\). Similarly for each \(j \in {\cal B}_{MT}\), \({\cal C}_j={\cal B}_{BS}\). Thus, for the downlink and uplink, we can get the optimal puDoF described in Sections \ref{sec:dl} and \ref{sec:ul} when $N_c < \frac{L}{2}$.
\linebreak
For the case where \(N_c \geq L + 1\), the coding scheme that achieves the inner bound in \eqref{eq:uldlzf2} is as follows. First, we associate each mobile terminal with the $L+1$ base stations connected to it. This achieves the puDoF value of unity during the uplink in the same way as the scheme that achieves it in Section \ref{sec:ul}. Hence, we know so far that \({\cal C}_{i} \supseteq \{i, i-1, i-2, \cdots, i - L\}\cap[K], \forall i\in[K]\).

We define ${\cal C}_i^D$ as the set of extra associations that the downlink scheme requires for MT $i$. Thus, $\forall i \in [K]$, we have that ${\cal C}_i = {\cal C}_i^D \cup \{i, i-1, \cdots, i - L\}$. 
For the downlink, we divide the network into disjoint subnetworks; each consists of \(L + 2\left(\left\lceil \frac{L + 1}{2} \right\rceil + N_c - (L + 1)\right)\) consecutive transmitter-receiver pairs. We define $\epsilon = \lceil\frac{L+1}{2}\rceil$, and $\kappa = \epsilon + N_c - (L+1)$. The cell association has a repeated pattern every $2\kappa + L$ BS-MT pairs, and hence, it suffices to describe it for the first $2\kappa+L$ BS-MT pairs. We consider two cases based on the parity of the connectivity parameter $L$. If L is odd, we partition the indices of mobile terminals in the subnetwork into three sets:
\begin{eqnarray*} 
	{\cal S}_1 &=& \{\epsilon, \epsilon + 1, \cdots, \epsilon + \kappa -1\},
	\\{\cal S}_2 &=& \{2\epsilon + \kappa, 2\epsilon + \kappa + 1 \cdots, 2\epsilon + 2\kappa -1\},
    \\{\cal S}_3 &=& \{1, 2, \cdots, L + 2\kappa\} \setminus ({\cal S}_1 \cup {\cal S}_2).
\end{eqnarray*}
The mobile terminals indexed in ${\cal S}_3$ are kept inactive. The cell associations for downlink are given by the following description.

${\cal C}_{i}^D=
\begin{cases}
\{1, 2, \cdots, \kappa\}, \quad &\forall i \in {\cal S}_1,\\
\{\epsilon + \kappa, \epsilon + \kappa + 1, \cdots, \epsilon + 2\kappa - 1\},\quad &\forall  i \in {\cal S}_2.
\end{cases}$\\

If L is even, we partition the indices of mobile terminals in the subnetwork into three sets:
\begin{eqnarray*} 
	{\cal S'}_1 &=& \{\epsilon, \epsilon + 1, \cdots, \epsilon + \kappa -1\},
	\\{\cal S'}_2 &=& \{2\epsilon + \kappa - 1, 2\epsilon + \kappa + 1 \cdots, 2\epsilon + 2\kappa -2\},
    \\{\cal S'}_3 &=& \{1, 2, \cdots, L + 2\kappa\} \setminus ({\cal S}_1 \cup {\cal S}_2).
\end{eqnarray*}

The mobile terminals indexed in ${\cal S}_3'$ are kept inactive. The downlink cell associations are given by the following description.

${\cal C}_{i}^D=
\begin{cases}
\{1, 2, \cdots, \kappa\}, \quad &\forall i \in {\cal S'}_1,\\
\{\epsilon + \kappa, \epsilon + \kappa + 1, \cdots, \epsilon + 2\kappa - 1\},\quad &\forall  i \in {\cal S'}_2.
\end{cases}$\\

Hence, if $L$ is odd, we have a subnetwork of $L + 2\kappa$ transmitter-receiver pairs, and we decode $$ (\epsilon + \kappa - 1 - \epsilon + 1) + (2\epsilon + 2\kappa -1 - (2\epsilon + \kappa) + 1)= 2\kappa$$ words during the downlink, and hence the puDoF during the downlink is $$\frac{2\kappa}{L + 2\kappa} = \frac{2\left(\frac{L+1}{2} + \left(N_c - (L+1)\right)\right)}{L + 2\left(\frac{L+1}{2} + \left(N_c - (L+1)\right)\right)}.$$ 

\begin{figure}
    \centering
    \includegraphics[width=0.4\columnwidth]{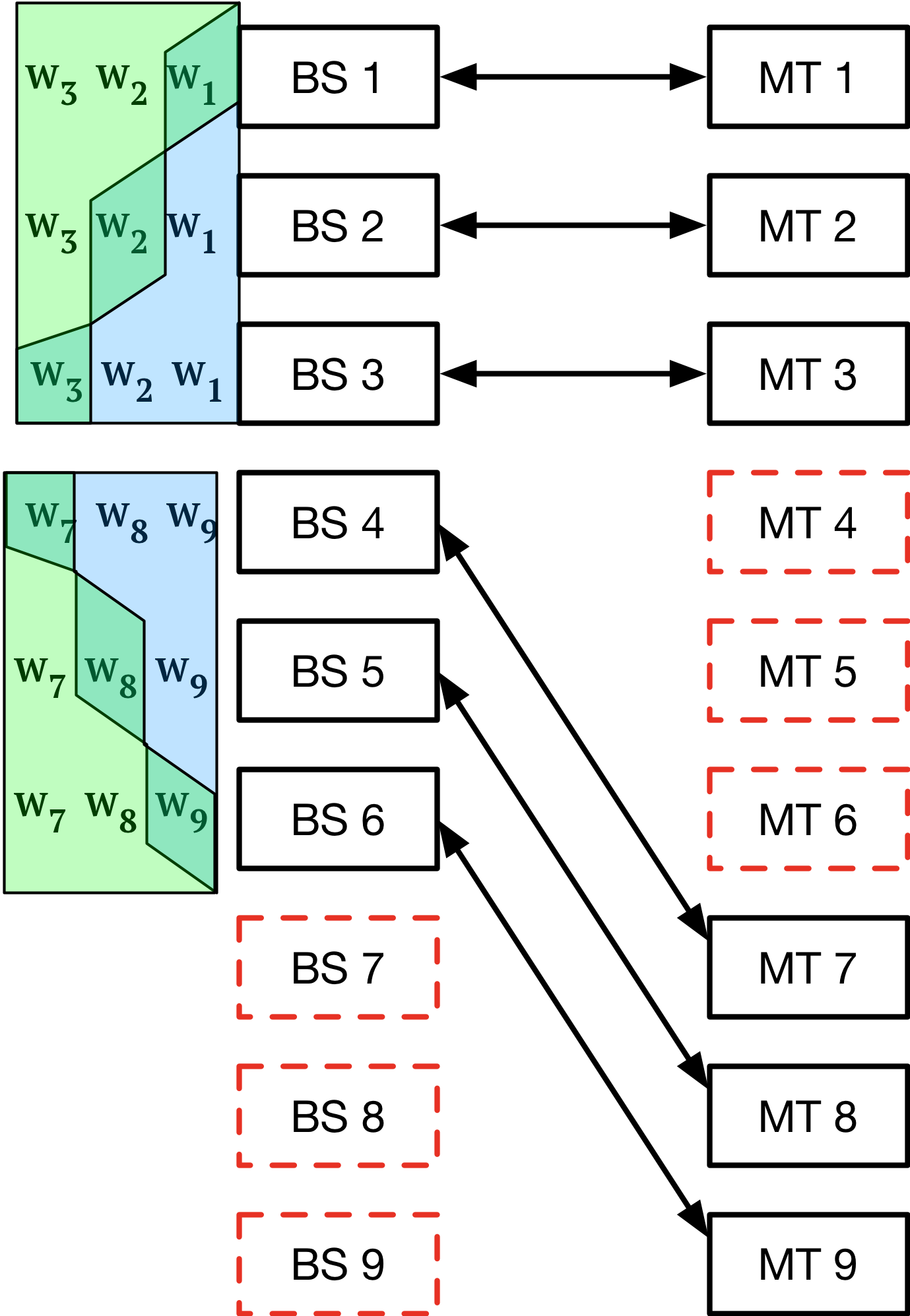}
    \caption{Scheme for average uplink (green shade) and downlink (blue shade) communication when \(N_c \leq L\), $N_c$ = 3, $L$ = 3}
    \label{fig:n_c_leq_L}
\end{figure}

A similar argument follows for the case when $L$ is even.  We have a subnetwork of $L + 2\kappa$ transmitter-receiver pairs and we decode $$ (\epsilon + \kappa - 1 - \epsilon + 1) + (2\epsilon + 2\kappa -2 - (2\epsilon + \kappa - 1) + 1)= 2\kappa$$ words during the downlink, which leads us to the same inner bound.
The proof of the inner bounds in \eqref{eq:uldlzf2} is hence complete.
Figures \ref{fig:n_c_leq_L} and \ref{fig:n_c_geq_L} serve as examples for the average uplink-downlink inner bounds defined in this section.

\begin{figure}[t!]
    \centering
    \subfloat[][L odd, $L = 3$, $N_c$ = 4]{\includegraphics[width=0.5\columnwidth]{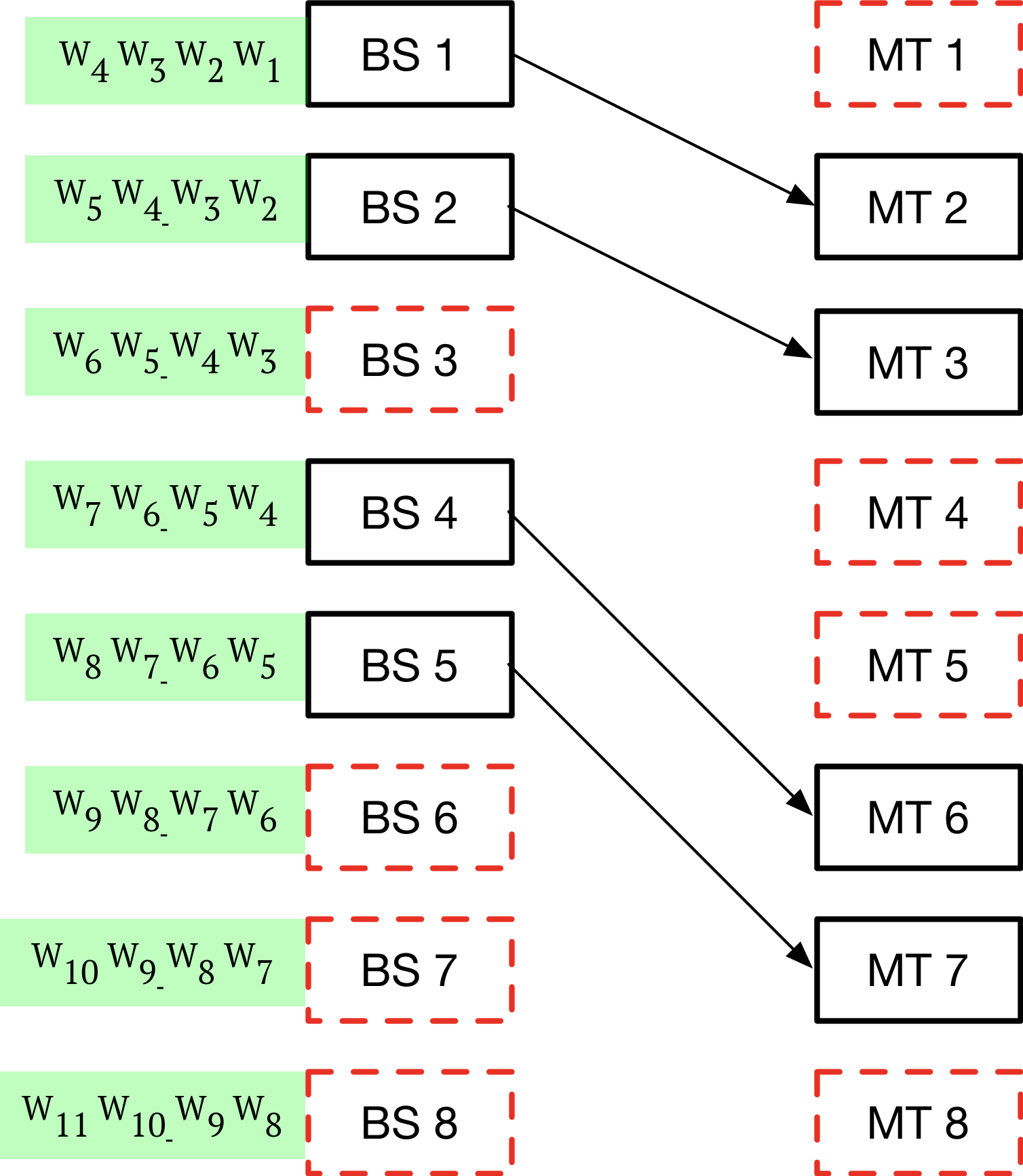}\label{fig:n_c_geq_L_1}}
     \subfloat[][L even $L = 2$, $N_c$ = 4]{\includegraphics[width=0.5\columnwidth]{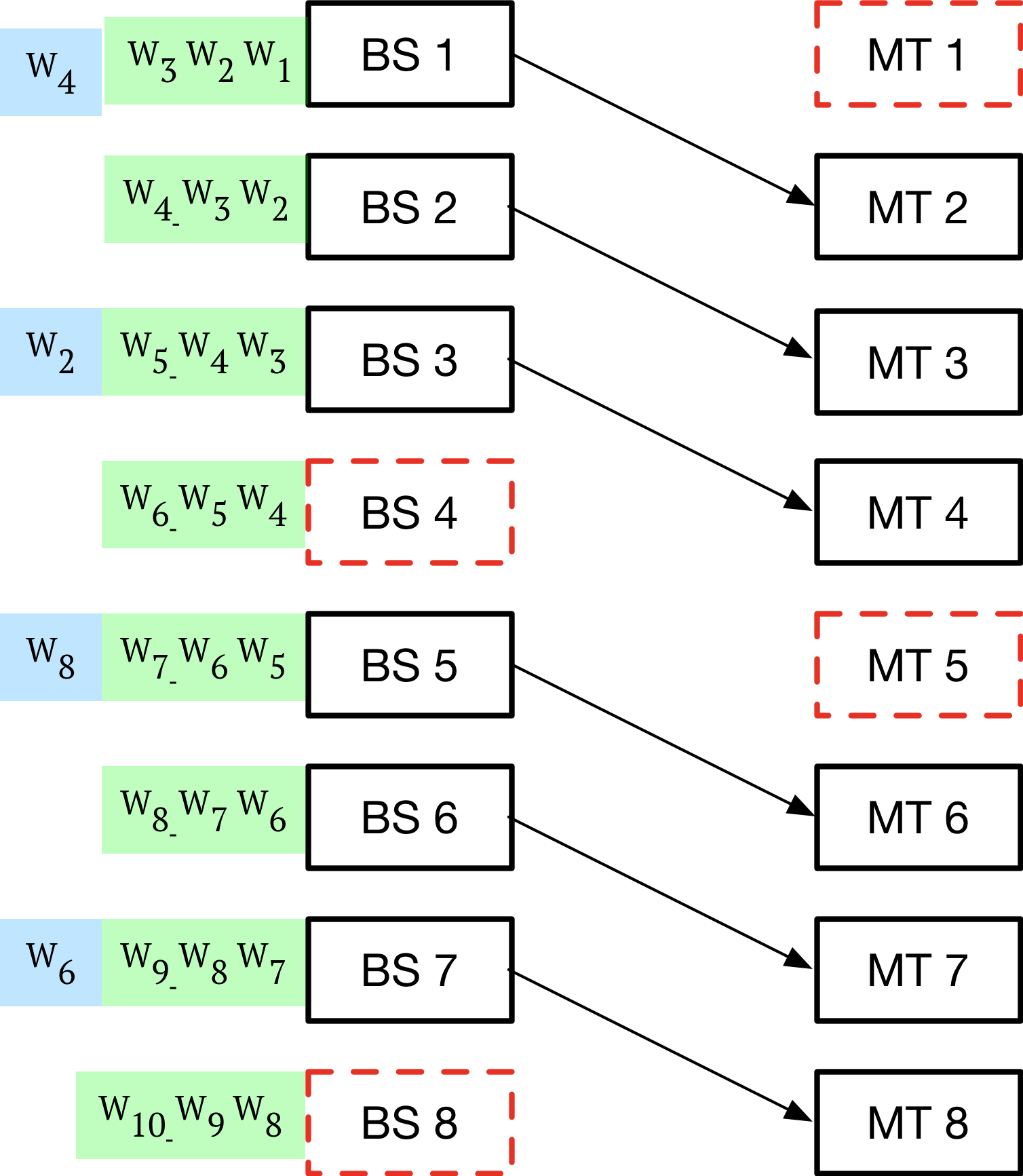}\label{fig:n_c_geq_L_2}}
    \caption{Scheme for downlink, with all the associations needed for optimal uplink, that achieves the lower bound defined in equation \eqref{eq:uldlzf2} when \(N_c \geq L + 1\)}
    \label{fig:n_c_geq_L}
\end{figure}


\subsection{Converse Proof for Full Coverage Associations}\label{sec:fc}
We show that the downlink puDoF as described in Theorem~\ref{thm:dlunityul} is optimal when we have unity DoF for the uplink, i.e., each mobile terminal is associated with all the base stations connected to it. In other words, we are restricted in this section to full coverage cell association schemes. 

We need to borrow the result on downlink cooperative zero-forcing in \cite[Lemma $2$]{Bande-ElGamal-Veeravalli-arXiv16} before proving the main result of this section, stated in Theorem~\ref{thm:dlunityul}. In order to state this result, we need to make the following definitions for any cooperative zero-forcing scheme. For any set ${\cal S} \subseteq [K]$, let ${\cal V}_{\cal S}$ be the set of active receivers connected to transmitters in ${\cal S}$. Also, for each transmitted message $W_i$, we define $\tilde{\cal C}_i \subseteq {\cal C}_i$ as the set of indices of base stations that are actively transmitting $W_i$. We finally say that there exists a matching between a set of base station transmitters and a set of mobile terminal receivers, if there is a matching between the vertices corresponding to these nodes in the bipartite interference graph.
\begin{lemma}[\cite{Bande-ElGamal-Veeravalli-arXiv16}]\label{lem:aux}
Using cooperative zero-forcing in the downlink, it has to be the case that for each transmitted message $W_i$, there exists a matching between transmitters in $\tilde{\cal C}_i$ and the set of active receivers connected to them ${\cal V}_{\tilde{\cal C}_i}$, and the matching covers all such active receivers. 
\end{lemma}
We now prove Theorem~\ref{thm:dlunityul}.
We define $n_j$ to be the number of active receivers with an index that is less than or equal to $j$. The key idea of the proof is to show the following for any downlink zero-forcing scheme with a full coverage association in a large network,
\begin{equation}\label{eq:aux}
\forall i\in {\bf Z}^+, n_{(2\kappa+L)i} \leq 2\kappa i.
\end{equation}
In other words, if we split the network into subnetworks, each consists of consecutive $2\kappa+L$ transmitter-receiver pairs, then the number of active receivers in each subnetwork is at most $2\kappa$. Without loss of generality, we restrict our attention to scenarios where $L$ is odd. Also, we assume that the first base station (BS $1$) is active, noting that otherwise we could establish our argument from the first active base station. We also assume that $|{\cal C}_i| \leq N_c-(L+1)+i, \forall i \in \{1,2,\cdots,L\}$, and note that this additional constraint cannot affect the puDoF value, since we are imposing it only for a fixed number of mobile terminals, regardless of the network size.

We prove \eqref{eq:aux} by induction. The base of the induction is the following,
\begin{equation}\label{eq:base}
n_{2\kappa+L}\leq 2\kappa,
\end{equation}
and if the bound is met tightly, then receivers $\{\kappa+L+1,\kappa+L+2,\cdots,2\kappa+L\}$ are active, and there are at most $\epsilon$ inactive receivers with indices in $\{\kappa+1,\kappa+2,\cdots,\kappa+L\}$. In other words, if there are $2\kappa$ active receivers in the first subnetwork, then the last $\kappa$ receivers in that subnetwork have to be active, and there are at most $\epsilon$ inactive receivers among the $L$ preceding receivers. The induction step would then be to show that for the $i^\textrm{th}$ subnetwork, if $n_{(2\kappa+L)(i-1)} = 2\kappa (i-1)$, and it is either the case that the last $\kappa$ receivers in subnetwork $i-1$ are active and the preceding $L$ receivers have at most $\epsilon$ inactive receivers, or it is the case that the last $\kappa-1$ receivers in subnetwork $i-1$ are active and the preceding $L$ receivers have at most $\epsilon-1$ inactive receivers, then it follows that $n_{(2\kappa+L)i} \leq 2\kappa i$, and if the bound is met tightly, then it is either the case that the last $\kappa$ receivers in subnetwork $i$ are active and the preceding $L$ receivers have at most $\epsilon$ inactive receivers, or it is the case that the last $\kappa-1$ receivers in subnetwork $i$ are active and the preceding $L$ receivers have at most $\epsilon-1$ inactive receivers. Note that for simplicity of the proof, we ignore the case when the bound in \eqref{eq:aux} is not met tightly for any value of $i$, as that case would follow in a straightforward manner from the body of the induction proof.

Consider the case when BS $1$ is transmitting to MT $1$, i.e., $I(W_1;X_1)>0$. Also, note that MT $1$ can only be associated with $N_c-(L+1)=\kappa-\epsilon$ base stations other than BS $1$. Let $x$ be the largest index of a base station actively transmitting $W_1$, then it follows from Lemma~\ref{lem:aux} that $n_{x+L} \leq \min(x,\kappa-\epsilon+1)$. Hence, we have that $n_{\kappa+\epsilon} \leq \kappa-\epsilon+1=\kappa+\epsilon-L$, and hence, the base statement would follow. We now show that the base statement holds even after relaxing the assumption that MT $1$ is active. Let $k \leq \epsilon$ be the smallest index of an active mobile terminal, and assume without loss of generality that BS $1$ is actively transmitting $W_k$. $W_k$ can only be associated with $N_c-(L+1)$ base stations with an index greater than $k$, and hence it follows from Lemma~\ref{lem:aux} that $|{\cal V}_{\tilde{\cal C}_k}| \leq N_c-(L+1)+k$. In other words, since $W_k$ can only be transmitted from the first $k$ base stations as well as other $N_c-(L+1)$ base stations, the number of active receivers connected to transmitters actively transmitting $W_k$ is at most $N_c-(L+1)+k=\kappa-\epsilon+k$. Let $x$ be the largest index of a base station transmitter actively transmitting $W_k$, then we know from Lemma \ref{lem:aux} that $n_{x+L} \leq \min (x,\kappa-\epsilon+k)$. It follows from the assumption that $k \leq \epsilon$ that $n_{\kappa+L} \leq \kappa$, and hence, the base statement would follow. If the smallest index of an active mobile terminal $k > \epsilon$, we only consider the case when $n_{\kappa+L} > \kappa$, as otherwise, the base statement would follow as in the above considered cases. Note that if $n_{\kappa+L} > \kappa$ and the first $\epsilon$ receivers are inactive, then it has to be the case that $\epsilon > 1$. Assume for simplicity that $n_{\kappa+L} = \kappa+1$, and let $m$ be the largest index of an active mobile terminal among the first $\kappa+L$. We now consider the following two possibilities:
\begin{itemize}
    \item If ${\cal V}_{\tilde{\cal C}_m}$ does not contain all the $\kappa$ active receivers with indices less than $m$, then it follows by inspection that all the transmitters actively transmitting $W_m$ have an index that is at least equal to $m-\epsilon+2$. Because $W_m$ can only be available at $\kappa-\epsilon$ transmitters outside its interference set, it would then follow from Lemma~\ref{lem:aux} that there are at most $\kappa-\epsilon+1$ active receivers among the last $\kappa+\epsilon-1$ receivers in the first subnetwork. In particular, there would be at most $\kappa-\epsilon$ active receivers in the first subnetwork with an index greater than $m$, and hence, $n_{2\kappa+L} \leq \kappa+1+\kappa-\epsilon < 2\kappa$.
    \item If ${\cal V}_{\tilde{\cal C}_m}$ contains all the $\kappa$ active receivers with indices less than $m$, then since we can only assign $W_m$ to at most $\kappa-\epsilon$ base stations outside its mobile terminal's interference set, it follows from Lemma~\ref{lem:aux} that at least $\epsilon+1$ transmitters in the interference set of MT $m$ are actively transmitting $W_m$. It would then follow that at least $\epsilon$ receivers succeeding MT $m$ are inactive, and hence, $n_{2\kappa+L} < 2\kappa$ in this case as well.
\end{itemize}
We conclude from all the above considered cases that the base statement holds. We now prove the induction step, and for simplicity consider the second subnetwork, i.e., the case when $i=2$. If no base station from the first subnetwork is actively transmitting a message that belongs to the second subnetwork, then the proof would be identical to the base proof above. It hence suffices to consider the cases when base stations from the first subnetwork are used. Let $k$ be the smallest index of an active mobile terminal in the second subnetwork, and let $x$ be the smallest index of a base station actively transmitting $W_k$. If $x \geq \kappa+L+1$, then because the last $\kappa$ receivers in the first subnetwork are active, it would follow from Lemma~\ref{lem:aux} that there would be a base station in the second subnetwork actively transmitting $W_k$, and the proof would follow as for the base case above. If $x_k < \kappa+L+1$, then $W_k$ would cause interference at all active $\kappa$ receivers at the end of the first subnetwork, and hence, at least $\epsilon+1$ base stations in the interference set of MT $k$ would actively transmit $W_k$. It then follows that if $k \geq \epsilon$, at least one base station in the second subnetwork would actively transmit $W_k$, and the proof would again follow as the base case above. If $k \leq \epsilon-1$, and no transmitter in the second subnetwork is actively transmitting $W_k$, then at least $\epsilon$ receivers succeeding $k$ would be inactive, because of interference caused by $W_k$. We could then show, using a similar argument as in the base case above, that $n_{4\kappa+2L}\leq 4\kappa$, and if the bound is met tightly then $k=1$, and the last $\kappa-1$ receivers in the second subnetwork are active, and the $L$ preceding receivers have at most $\epsilon-1$ inactive receivers. The proof of the induction step for $i \geq 3$ would then be similar to that of $i=2$. 

The induction proof is hence complete, and it implies that \eqref{eq:aux} holds for all positive integer values of $i$, and the hence, the theorem statement holds.

\subsection{Converse Proof for Wyner's Linear Network ($L=1$)}\label{sec:lone}
In this section, we show that for $L = 1$, the lower bound of Theorem~\ref{thm:avguldl} is information-theoretically optimal. More precisely, we prove Theorem~\ref{thm:lone}.
The proof of achievability for Theorem~\ref{thm:lone} follows from the inner bound proof in Section~\ref{sec:uldl}. For the case where $N_c=1$, the upper bound follows from the fact that the maximum per user DoF for each of the downlink and uplink sessions is $\frac{2}{3}$, even if we are allowed to change the cell association between the uplink and downlink. The proof of the downlink case is provided in~\cite{ElGamal-Annapureddy-Veeravalli-IT14}, and the proof of the uplink case is similar. 


Before making the main argument, we first need the following auxiliary lemmas for finding a converse for the uplink scenario.
\begin{lemma}\label{lem:uplinklemone}
Given any cell association and any coding scheme for the uplink, the per user DoF cannot be increased by adding an extra association of mobile terminal $i$ to base station $j$, where $j \notin \{i,i-1\}$.
\end{lemma}
\begin{IEEEproof}
The lemma states that associating any mobile terminal to a base station that is not connected to it cannot be \emph{useful} for the uplink case. The key fact validating this lemma is that unlike the downlink case, the knowledge of a message at a base station cannot allow for the possibility of propagating the interference caused by this message beyond the two original receivers that are connected to the transmitter responsible for delivering the message. In other words, no matter what cell association we use for mobile terminal $i$, the message $W_i$ will not cause interference at any base station except base stations $i$ and $i-1$, and hence, having this message at any other base station cannot help neither in decoding the message nor in canceling interference. In what follows, we detail the formal argument.

Given any cell association scheme, assume we have a reliable communication scheme with block length $n$, where the decoder at each receiver with index $k$ uses the signal $\hat{Y}_k^n=f_k(Y_k^n,\{W_i: k\in{\cal C}_i\})$ to obtain an estimate of $W_k$. The signal $\hat{Y}_k^n$ is obtained using a - possibly random - function $f_k$ from the received signal $Y_k^n$, as well as side information about all the messages associated with BS $k$. We show that under this assumption, one can always construct a reliable communication scheme, where the decoder at each receiver with index $k$ uses a signal $\tilde{Y}_k^n=\tilde{f}_k(Y_k^n, \{W_i: k\in{\cal C}_i\cap\{i-1,i\}\})$ to obtain an estimate of $W_k$. The signal $\tilde{Y}_k^n$ is obtained using a function $\tilde{f}_k$ from the received signal $Y_k^n$, as well as side information about all the messages whose mobile terminal is connected to BS $k$ and are associated with BS $k$. For each message $W_i$, we construct an independent random variable $Q_{i}$ that is \emph{stochastically equivalent} to $W_i$, i.e., $Q_i$ has the same alphabet and distribution as $W_i$. We then let,
\begin{equation*}
    \tilde{f}_k(Y_k^n, \{W_i: k\in{\cal C}_i\cap\{i-1,i\}\})=f_k\left(Y_k^n,\{W_i: k\in{\cal C}_i\cap\{i-1,i\}\},\{Q_i: k\in{\cal C}_i, k\notin\{i-1,i\}\}\right).
\end{equation*}
Let $R_k$ be the rate achieved for user $k$ in the assumed reliable communication scheme.
We now observe that the following holds. 
\begin{eqnarray}
n\sum_k R_k &=& \sum_k \textsf{H}\left(W_k\right) 
\\&\overset{(a)}{\leq}&  \sum_k I\left(W_k;\{\hat{Y}_i^n: i\in[K], k\in{\cal C}_i\cap\{i-1,i\}\}\right)+ o(n) \label{eq1}
\\&\overset{(b)}{=}& \sum_k I\left(W_k;\{\tilde{Y}_i^n: i\in[K], k\in{\cal C}_i\cap\{i-1,i\}\}\right)+ o(n) \label{eq2}
\\&=& \sum_k \textsf{H}\left(W_k\right) - \textsf{H}\left(W_k|\{\tilde{Y}_i^n: i\in[K], k\in{\cal C}_i\cap\{i-1,i\}\}\right) + o(n),
\end{eqnarray}
where $\textsf{H}(.)$ is the entropy function for discrete random variables, and $(a)$ follows from Fano's inequality and the above assumption that the assumed reliable communication scheme uses the signals $\hat{Y}_k$ for decoding, as well as the fact that only received signals corresponding to base stations that are associated with and connected to a message's mobile terminal can be used for decoding the message. Also, $(b)$ holds because for each $i,k\in[K]$ such that $k\in{\cal C}_i$ and $k\notin\{i-1,i\}$, the received signal $Y_k$ is independent of the message $W_i$, and hence replacing $W_i$ with $Q_i$ leaves the joint distribution of the involved random variables in the mutual information expression of~\eqref{eq2} identical to that of~\eqref{eq1}. Now, it follows that,
\begin{equation}
\sum_k \textsf{H}\left(W_k|\{\tilde{Y}_i^n: i\in[K], k\in{\cal C}_i\cap\{i-1,i\}\right) = o(n),
\end{equation}
and hence, the rates $R_k, k\in[K],$ are achievable in the constructed scheme.

\end{IEEEproof}

Lemma~\ref{lem:uplinklemone} gives us two possibilities for choosing the cell association of mobile terminal $i$; either we associate it with both base stations $i$ and $i-1$ or only one of these base stations. We use Lemma~\ref{lem:uplinklemtwo} to upper bound the degrees of freedom for the latter case. First, we will need in the remainder of the proof to use the following generalization of~\cite[Lemma $4$]{ElGamal-Annapureddy-Veeravalli-IT14}. For any set ${\cal A} \subseteq [K]$, we define ${\cal U}_{\cal A}$ as the set of indices of transmitters that exclusively carry messages with indices in ${\cal A}$, and hence, the complement set $\bar{\cal U}_{\cal A}$ is the set of indices of transmitters that carry any message with indices outside ${\cal A}$.
\begin{lemma}(\cite{ElGamal-Annapureddy-Veeravalli-IT14})\label{lem:uplinkaux}
In either downlink or uplink sessions, if there exists a set ${\cal A}$ of messages that are decodable using a set of received signals $Y_{\cal B}$, a function $f_1$, and a function $f_2$ whose definition does not depend on the transmit power $P$, and $f_1(Y_{\cal B},X_{{\cal U}_{\cal A}})=X_{\bar{\cal U}_{\cal A}}+f_2(Z_{\cal B})$, then the sum DoF is bounded by the number of received signals in $Y_{\cal B}$. More precisely, $\eta \leq |{\cal B}|$.
\end{lemma}
\begin{IEEEproof}
The proof is almost identical to the proof of~\cite[Lemma $4$]{ElGamal-Annapureddy-Veeravalli-IT14} with appropriate change of variables, and hence we only provide a sketch here for brevity. Assuming a reliable communication scheme, if we are given the received signals $Y_{\cal B}$, messages $W_{\cal A}$ can be decoded reliably, and hence, the transmit signals $X_{{\cal U}_{\cal A}}$ can be reconstructed. If we can reconstruct the remaining transmit signals $X_{\bar{\cal U}_{\cal A}}$, then all messages could be decoded. From the hypothesis of the statement of the lemma, we know that the uncertainty in reconstructing the remaining transmit signals is due to Gaussian noise, which does not affect the degrees of freedom. The sum DoF is hence bounded by the number of received signals used for decoding all messages $|{\cal B}|$. 
\end{IEEEproof}
We note that in the downlink, the set ${\cal B}={\cal A}$, and $\bar{\cal U}_{\cal A}=\cup_{i \notin {\cal A}} {\cal C}_i$. In the uplink, ${\cal U}_{\cal A}={\cal A}$.
\begin{lemma}\label{lem:uplinklemtwo}
If either mobile terminal $i$ or mobile terminal $i+1$ is not associated with base station $i$, i.e., the following holds,
\begin{equation}
    |{\cal C}_i\cap\{i\}|+|{\cal C}_{i+1}\cap\{i\}|\leq 1,
\end{equation}
then it is either the case that the received signal $Y_i$ can be ignored in the uplink without affecting the sum rate, or it is the case that the uplink sum DoF for messages $W_i$ and $W_{i+1}$ is at most one, i.e., $d_i+d_{i+1} \leq 1$.
\end{lemma}
\begin{IEEEproof}
If neither $W_i$ nor $W_{i+1}$ is associated with base station $i$, then it is clear that $Y_i$ can be ignored in the uplink. Further if only one of the two message is associated with base station $i$ but is not decodable from $Y_i$ in the uplink, then we also can ignore this received signal. We now focus on the remaining case when exactly one of $W_i$ and $W_{i+1}$ is associated with base station $i$ and can be successfully decoded from $Y_i$ in the uplink. We assume without loss in generality that $W_i$ is the message associated with base station $i$.
We now create a new network identical to the original but with forcing all messages in the network other than $W_i$ and $W_{i+1}$ to be deterministic, and hence we have that the sum DoF $\eta = d_i+d_{i+1}$. We also note that $d_i$ and $d_{i+1}$ can be only be increased in the new setting, and hence if we obtain an upper bound on their sum, it would apply to the original values. We then apply Lemma~\ref{lem:uplinkaux} with ${\cal A}={\cal B}=\{i\}$ and obtain that $d_i+d_{i+1} \leq 1$.

\end{IEEEproof}

We first consider the case where each mobile terminal can be associated with two base stations, i.e., $N_c=2$. Fix a cell association scheme and divide the indices of the network into sets (subnetworks); each consists of consecutive three indices. We define $x$ to be the fraction of subnetworks, whose middle base station is only associated with at most one of the mobile terminals that are connected to it. We show that the uplink puDoF is at most $(1-x)+\frac{5}{6}x$, and the added puDoF due to downlink transmission is at most $\frac{2}{3}(1-x)+\frac{5}{6}x$, and hence, it would follow that $\tau(L=1,N_c=2) \leq \frac{5}{6}$ as stated in~\eqref{eq:lone}. We first show the uplink part. For each subnetwork whose middle mobile terminal is not associated with the two base stations connected it, Lemma~\ref{lem:uplinklemtwo} will apply for at least one of these two base stations set as base station $i$; let $x_1$ be the fraction of such subnetworks, where Lemma~\ref{lem:uplinklemtwo} implies that $d_i+d_{i+1} \leq 1$, and $x_2$ be the fraction of such subnetworks, where Lemma~\ref{lem:uplinklemtwo} implies that $Y_i$ can be ignored in the uplink. We have that $x=x_1+x_2$. We also have that the uplink puDoF is at most $1-\frac{1}{3}x_1$, since in each subnetwork counting towards $x_1$, at most $2$ DoF are achieved for the three users of the subnetwork in the uplink. Also, the uplink puDoF is at most $1-\frac{1}{3}x_2$, since at least $x_2$ received signals are ignored in the uplink. It follows that the uplink puDoF is at most $1-\frac{1}{3}\max(x_1,x_2)$. It hence follows that it is at most $1-\frac{1}{6}x$. We now bound the added puDoF due to downlink transmission. For each subnetwork whose middle mobile terminal $i$ is associated with the two base stations connected to it, we apply Lemma~\ref{lem:uplinklemone} within the subnetwork with the set ${\cal A}=\{i-1,i+1\}$ to conclude that at most $2$ DoF can be achieved for the three users in the subnetwork (note that $\{i+1\} \subseteq {\cal U}_{\cal A}$), and hence we lose at least $\frac{1}{3}(1-x)$ per user DoF. It hence suffices to show that in addition to that, we have to lose at least $\frac{1}{6}x$ per user DoF. Let ${\cal S}$ be the superset whose elements are sets of three indices each, representing subnetworks whose middle mobile terminal is associated with at most one of the base stations connected to it. It suffices to show that we have to lose at least $\frac{|{\cal S}|}{2}$ DoF in a large network. The proof is based on the following two facts on upper bounding the downlink DoF:
\begin{itemize}
    \item {\bf Fact 1:} For every five messages with consecutive indices, the achieved DoF is at most $4$. This follows by applying the irreducible message assignment lemma of~\cite{ElGamal-Annapureddy-Veeravalli-IT14} to the middle message, and then applying Lemma~\ref{lem:uplinklemone} with the set ${\cal A}$ consisting of all five indices except the middle index.
    \item {\bf Fact 2:} For every three messages with consecutive indices, if the middle message is associated with both base stations connected to its mobile terminal, then the achieved DoF is at most $2$. This follows by applying Lemma~\ref{lem:uplinklemone} with the set ${\cal A}$ consisting of all three indices except the middle index.
\end{itemize}
Consider the partitioning of ${\cal S}$ that puts every maximal set of subnetworks with consecutive indices in one partition. For any subset ${\cal P} \subseteq {\cal S}$ representing a partition, if it has an even number of elements or an odd number that is greater than $3$, then Fact $1$ would imply that we have to lose at least $\frac{|{\cal P}|}{2}$ DoF within that partition. We can hence restrict our attention to paritions containing $1$ or $3$ elements. If $|{\cal P}|=3$, note that we have to lose at least $1$ DoF among the first five messages included in the first two subnetworks in the partition because of Fact $1$. Further, due to the same fact, there is a DoF lost among the five messages consisting of the last four messages in ${\cal P}$ and the succeeding message which lies at the top of a subnetwork - not in ${\cal S}$ - that we upper bounded its uplink DoF by $3$ and downlink DoF by $2$; call this subnetwork $\tilde{s}$. If the following subnetwork to $\tilde{s}$ is in ${\cal S}$, then we bound the DoF of the five messages consisting of the last two in $\tilde{s}$ and the three of the following subnetwork using Fact $1$. It hence follows in this case that we have to lose at least an extra DoF, and hence, we have to lose more than $\frac{{|{\cal P}|}}{2}$ extra DoF that were not considered before. If the following subnetwork to $\tilde{s}$ is also not in ${\cal S}$, then consider the set of consecutive subnetworks consisting of $\tilde{s}$ and all succeeding subnetworks that are not in ${\cal S}$. We note that it is either the case that each mobile terminal, except the first, in the considered set of subnetworks is associated with the two base stations connected to it, or it follows from Lemma~\ref{lem:uplinklemtwo} that we lose at least $\frac{1}{2}$ DoF in the uplink due to associations in these subnetworks (using a similar argument to the uplink upper bound above), and hence, we lose overall $\frac{|{\cal P}|}{2}$ DoF that were not considered before. If it is the former, then we know using Fact $2$ above that we lose $1$ DoF among the three messages consisting of the second and third in $\tilde{s}$ and the first in the following subnetwork. We can then repeatedly apply Fact $2$ among the three messages consisting of the second and third in the current subnetwork and the first in the following subnetwork, as long as the following subnetwork is not in ${\cal S}$. If the following subnetwork is in ${\cal S}$, then we use Fact $1$ to imply that we lose an extra DoF among the five messages consisting of the second and third in the current subnetwork, and the three of the following subnetwork, and hence, we lose in this case $\frac{|{\cal P}|+1}{2}$ DoF due to the subnetworks in ${\cal P}$ as well as the first subnetwork in the next set in the partition. In this last case, we restart the argument from the second subnetwork in the next set of the partition, instead of the first; we remove the details of this step for brevity, as it is a very similar argument to the considered one. We hence have shown that if $|{\cal P}|=3$, then we lose at least $\frac{1}{2}$ puDoF by considering all subnetworks in ${\cal P}$. It hence remains to consider the case when $|{\cal P}|=1$. In this case, we use Fact $1$ to bound the DoF of the five messages consisting of the three in the subnetwork of ${\cal P}$ and the last in the preceding subnetwork and the first in the succeeding subnetwork. The proof then follows in a similar fashion to that for the case when $|{\cal P}|=3$, but by considering both preceding and succeeding subnetworks, instead of only succeeding subnetworks. The key idea is that a DoF bound that includes a message, other than the middle one, in a subnetwork not in ${\cal S}$ results in a DoF loss, either in uplink or downlink, of at least $\frac{1}{2}$. 


The extension of the above argument for $N_c > 2$ is straightforward, and hence, we omit it here for brevity. The main argument would rely on subnetworks; each consisting of $2N_c-1$ users, and using the same definition of $x$ as above, one can show that the uplink puDoF is at most $(1-x)+\frac{4N_c-3}{4N_c-2}x$, and the added puDoF due to downlink is at most $\frac{2N_c-2}{2N_c-1}(1-x)+\frac{4N_c-3}{4N_c-2}x$, and hence, it would follow that $\tau(L=1,N_c) \leq \frac{4N_c-3}{4N_c-2}$ as stated in \eqref{eq:lone}. The network will be split into subnetworks; each of size $2N_c-1$, and the uplink DoF upper bounding argument would remain the same as for $N_c=2$, because of Lemma~\ref{lem:uplinklemone}. The downlink argument will also be very similar to the case where $N_c=2$, but with replacing Fact $1$ above to imply a bound on the DoF of $2N_c$ for every consecutive $2N_c+1$ messages, and replacing Fact $2$ above to imply a bound of $2N_c-2$ DoF for every consecutive $2N_c-1$ messages whose middle message has a full coverage associations.  

\section{Discussion}\label{sec:discussion}
\subsection{When Separate Uplink-Downlink Optimization is Sub-optimal}
One important insight we observe from the results we obtained in this work, is that when the cell association constraint is small enough with respect to the connectivity parameter $\left(N_c \leq \frac{L}{2}\right)$, then the average zero-forcing puDoF is identical to that of either the downlink or uplink. In other words, there is no loss in this case due to making the cell association decisions based on the optimization for either downlink or uplink sessions. However, it is worth noting that zero-forcing is strictly sub-optimal - from an information theoretic standpoint - when $N_c < \frac{L}{2}$, as non-cooperative asymptotic interference alignment can lead to achieving $\frac{1}{2}$ puDoF in either downlink or uplink. Further, for higher values of $N_c$, it is obvious from our results that there is a tradeoff between optimizing the cell associations for the downlink or uplink.   

\subsection{Association Strategy for General Network Models}
We also note that when $\frac{L}{2} < N_c \leq L$, the proposed scheme for maximizing the average zero-forcing puDoF also leads to achieving the downlink-optimal zero-forcing puDoF, while for the case when $N_c > L$, it leads to achieving the uplink-optimal zero-forcing puDoF. While it may be intuitive to think that the latter observation would hold for more general network models, it is worth investigating whether the former observation would. More specifically, would it always be the case that for very low values of $N_c$, the average-optimal zero-forcing scheme would achieve optimal zero-forcing puDoF values for each of the uplink and downlink, and then for slightly higher values of $N_c$, the downlink-optimal zero-forcing puDoF is achieved, and then for higher values of $N_c$, the uplink-optimal zero-forcing puDoF is achieved?
We believe the answer to this question is yes for the considered locally connected network models, as we conjecture that the proposed scheme for maximizing the average zero-forcing puDoF is indeed optimal.  

\subsection{Interference Propagation and its Impact on Converse Proofs}
Finally, it is worth noting the distinction between the zero-forcing puDoF upper bound arguments that we had for the downlink and uplink. In the downlink, the argument we presented in Section~\ref{sec:fc} as well as the argument used in~\cite{ElGamal-Annapureddy-Veeravalli-IT14} to prove the result summarized in Section~\ref{sec:dl} rely on bounding the puDoF achieved in each subnetwork by the desired bound that applies for the overall puDoF. On the contrary, the arguments used in the uplink in Section~\ref{sec:ul} cannot follow the same footsteps, as it is possible to exceed the puDoF bound for certain subnetworks, at the cost of failing to meet it in neighboring subnetworks, due to the effect of borrowing or blocking base stations across subnetworks. The key reason underlying this difference is that when a message is shared over the backhaul for cooperative zero-forcing in the downlink, it causes interference at more mobile terminal receivers. On the other hand, when a message is shared over the backhaul for zero-forcing decoding over the uplink, its interference does not propagate to other base station receivers. This added restriction due to interference propagation in the downlink allows us to simplify the zero-forcing upper bound proofs by considering only subnetwork-only decoding.      

\section{Conclusion}\label{sec:conclusion}

In this work, we presented an effort to understand optimal cell association decisions in locally connected interference networks, focusing on optimizing for the average uplink-downlink puDoF problem. We considered a backhaul constraint that allows for associating each mobile terminal with $N_c$ base stations (cells), and an interference network where each base station is connected to a corresponding mobile terminal as well as $L$ mobile terminals with succeeding indices. We characterized the optimal cell association and puDoF for the uplink problem when zero-forcing schemes are considered. We also found that the characterization of the optimal association for the average uplink-downlink puDoF problem when $N_c \leq \frac{L}{2}$ follows from our uplink characterization and previous work for the downlink problem. We also presented the optimal zero-forcing downlink scheme if we fix the uplink scheme to the uplink-only-optimal scheme when $N_c \geq L+1$. We conjecture that it is in fact optimal to have full DoF in the uplink when $N_c \geq L+1$, and hence it would follow that the presented cell association and average puDoF are optimal in this case. Finally, we presented general inner bounds for the zero-forcing average puDoF across both uplink and downlink, and showed that they are information-theoretically optimal for Wyner's linear network, i.e., when $L=1$. For future work, we plan to consider validating the insights obtained through the results in this work to more general and closer-to-practice cellular network models.


\bibliography{refs}\label{refs}

\bibliographystyle{IEEEtran}

\end{document}